\documentclass[fleqn,usenatbib]{mnras}

\usepackage{newtxtext,newtxmath}

\usepackage[T1]{fontenc}

\DeclareRobustCommand{\VAN}[3]{#2}
\let\VANthebibliography\thebibliography
\def\thebibliography{\DeclareRobustCommand{\VAN}[3]{##3}\VANthebibliography}

\usepackage{graphicx}	
\usepackage{amsmath}	

\usepackage{amssymb}	
\usepackage[table,xcdraw]{xcolor}
\usepackage{physics}
\usepackage{tikz}
\usepackage{mathdots}
\usepackage{cancel}
\usepackage{color}
\usepackage{siunitx}
\usepackage{array}
\usepackage{multirow}
\usepackage{gensymb}
\usepackage{tabularx}
\usepackage{extarrows}
\usepackage{booktabs}
\usepackage{subfigure}
\usepackage{titlesec}
\usepackage{changepage}
\usetikzlibrary{fadings}
\usetikzlibrary{patterns}
\usetikzlibrary{shadows.blur}
\usetikzlibrary{shapes}

\titleformat{\paragraph}
{\normalfont\normalsize}{\theparagraph}{1em}{}
\titlespacing*{\paragraph}
{0pt}{3.25ex plus 1ex minus .2ex}{1.5ex plus .2ex}

\title[Tidal Dissipation in Binary-Single Encounters]{Influence of tidal dissipation on outcomes of  binary-single encounters between stars and black holes in stellar clusters}

\author[L. Hellström et al.]{
Lucas Hellström$^{1}$\thanks{E-mail: hellstrom@camk.edu.pl}, Abbas Askar$^2$, Alessandro A. Trani$^{3,4}$, Mirek Giersz$^1$, Ross P. Church$^2$ and Johan Samsing$^{5}$
\\
$^{1}$Nicolaus Copernicus Astronomical Centre, Polish Academy of Sciences, Warsaw, Poland\\
$^{2}$Lund Observatory, Department of Astronomy, and Theoretical Physics, Lund University, Box
43, SE-221 00 Lund, Sweden\\
$^{3}$Department of Earth Science and Astronomy, College of Arts and Sciences, The University of Tokyo, 3-8-1 Komaba, Meguro-ku, Tokyo 153-8902, Japan\\
$^{4}$Okinawa Institute of Science and Technology, 1919-1 Tancha, Onna-son, Okinawa 904-0495, Japan\\
$^{5}$Niels Bohr International Academy, Niels Bohr Institute, Blegdamsvej 17, 2100 Copenhagen, Denmark\\
}

\date{Accepted XXX. Received YYY; in original form ZZZ}

\pubyear{2022}

\begin{document}
\label{firstpage}
\pagerange{\pageref{firstpage}--\pageref{lastpage}}
\maketitle

\begin{abstract}
In the cores of dense stellar clusters, close gravitational encounters between binary and single stars can frequently occur. Using the \textsc{tsunami}  code, we computed the outcome of a large number of binary-single interactions involving two black holes (BHs) and a star to check how the inclusion of orbital energy losses due to tidal dissipation can change the outcome of these chaotic interactions. Each interaction was first simulated without any dissipative processes and then we systematically added orbital energy losses due to gravitational wave emission (using post-Newtonian (PN) corrections) and dynamical tides and recomputed the interactions. We find that the inclusion of tides increases the number of BH-star mergers by up to 75 per cent but it does not affect the number of BH-BH mergers. These results highlight the importance of including orbital energy dissipation due to dynamical tides during few-body encounters and evolution of close binary systems within stellar cluster simulations. Consistent with previous studies, we find that the inclusion of PN terms increases the number of BH-BH mergers during binary-single encounters. However, BH-star mergers are largely unaffected by the inclusion of these terms.

\end{abstract}

\begin{keywords}
gravitation -- methods: numerical -- stars: kinematics and dynamics -- binaries: close -- stars: black holes -- galaxies: star clusters: general
\end{keywords}



\section{Introduction}



Binary-single scatterings can frequently occur in the cores of dense globular clusters. These strong dynamical interactions result in cluster heating and are an important formation channel for stellar exotica and compact object binary systems \citep{mcmillan86,sigurdsson93,davies93}. There is a need to properly treat and determine the outcome of these interactions within stellar cluster evolution simulations. However, the task of analytically predicting the motions of three bodies in a closed system has been unsolved for hundreds of years \citep{poincare1892}. The problem stems from the lack of a closed-form solution to the equations of motion. In the absence of an analytic solution to this problem, several recent works have tried to statistically describe and predict the outcome of these interactions \citep{stone2019,kol2021,manwadkar2021,ginat2021,2021arXiv210806335P}. Commonly used direct \textit{N}-body and Monte Carlo simulation codes for evolving stellar clusters numerically integrate the equations of motions to determine the outcome of these chaotic interactions.

Dense stellar clusters could retain a sizeable fraction of stellar-mass black holes (BHs) that form when massive stars evolve and end their lives. \citep{morscher2015,askar2018}. Due to dynamical friction, these BHs can segregate to the centre of the stellar cluster where they can interact with each other and surrounding stars \citep{arcasedda2018,kremer2019b}. This can result in numerous binary-single encounters involving multiple BHs and/or stars in the cores of stellar clusters. These dynamical interactions can lead to the formation, hardening and ejections of binary BHs that could potentially merge due to gravitational wave radiation \citep[e.g.,][]{spz2000,moody2009,downing2010,tanikawa2013,bae2014,ziosi2014}. Therefore, the dynamical production of merging binary BHs in stellar clusters \citep[e.g.,][]{rod2016,askar2017,dicarlo2019,banerjee2021,rod2022} could be one of the channels for forming gravitational wave sources being observed with the ground-based LIGO/Virgo detectors \citep{ligo2016,ligo2021a,ligo2021b}. 

Recent studies examining merging BHs in dense environments have shown the importance of including orbital energy and angular momentum dissipation through gravitational waves during single-single \citep{samsing2020}, binary-single \citep{samsing2014,samsing2018} and binary-binary \citep{zevin2019,sedda2021} encounters in stellar clusters. The inclusion of these dissipative effects increases the number of merging binary BHs in stellar clusters and can also lead to eccentric BH mergers during few-body encounters involving BHs \citep{eccentricBHMergersDuringBinSingle,rodriguez2018a,rodriguez2018b,banerjee2018}. Recently \citet{2022Natur.603..237S} have shown that dynamical single and binary BH encounters in disks of active galactic nuclei can also produced eccentric BH mergers. The detection of these eccentric mergers would be useful in constraining the astrophysical origin of binary BHs \citep{samsing2017,LISA3,romeroshaw2020,zevin2021}. In order to include dissipative effects of gravitational wave emission in the numerical computation of few-body interactions, post-Newtonian (PN) correction terms \citep[e.g.][]{PN_fack, PN_Blanchet} to the Newtonian equations of motion are used \citep{kupi2006,harfst2008,mikkola2008}.

Scattering codes such as \textsc{Fewbody} by \citep{Fewbody_paper} and \textsc{tsunami}  \citep{tsunami_first, tsunami2} have been developed to compute the outcome of dynamical interactions for a small number of objects. These integrators solve the N-body equations of motion to advance the system forward in time. The \textsc{Fewbody} code has been used to compute the outcome of binary-single and binary-binary interactions within Monte Carlo simulation codes for stellar clusters such as \textsc{MOCCA} \citep{MOCCA1} and \textsc{CMC} \citep{rod2022}. There has been a lot of focus on including PN correction terms (up to 2.5PN) to capture the dissipative effects of gravitational wave emission when computing few-body interactions involving multiple BHs within stellar cluster simulations \citep[e.g.,][]{rodriguez2018a,rodriguez2018b,banerjee2018}. However, few studies have considered the effect of dissipation of tidal energy in few-body interactions involving BHs and stars \citep{ginat2021}.

Dynamical interactions between BHs and stars can frequently occur within stellar clusters. Such interactions can lead to the tidal disruption of stars \citep{fabian75,kremer2019a,fragione2019,kremer2022}. Binary-single interactions involving two BHs and a star are particularly interesting since the tidal disruption of stars by binary BHs may influence the properties of merging BHs \citep{lopez2019}. Given that the inclusion of PN corrections to the equations of motion when computing the outcome of binary-single encounters between BHs leads to an increase in their merger rate. In this paper, we investigate how the inclusion of tidal dissipation can influence the outcome of binary-single encounters involving two BHs and a star. For this purpose, we use the \textsc{tsunami}  code (see Section \ref{sec:simulations}) to carry out a few 100,000 scattering experiments to systematically check how the inclusion of PN terms and tides can affect the outcome of binary-single encounters between two BHs and a star. In order to do this, each encounter was first computed without the inclusion of any dissipative effects, we then computed the same encounter with the inclusion of PN corrections. After this the encounter was recomputed with inclusion of tides and then finally with both PN corrections and tides. We find that the inclusion of tidal dissipation results in a significant increase of BH-star mergers. We show the statistical significance of this result and point out specific examples where the inclusion of tidal effects results in a different outcome. These results point out towards the importance of including the effects of tidal dissipation during few-body encounters in evolving stellar clusters. 

In Section \ref{sec:method}, we describe the \textsc{tsunami}  code and its main features as well as a brief explanation of how the interactions are setup. In Section \ref{sec:testSetup}, we then describe the initial setup and results of the few-body encounters that we used to test the \textsc{tsunami} code. In Section \ref{sec:moccaSetup}, we describe and show results from the initial data set for binary-single encounters that we took from the \textsc{MOCCA}-Survey Database I \citep{askar2017}. The Section also contains specific examples which highlight the importance of including tidal dissipation effects during these encounters.

\section{Methods}\label{sec:method}

\subsection{The \textsc{tsunami} few-body code}
\label{sec:simulations}
We use the direct few-body integrator \textsc{tsunami}  \citep{tsunami_first, tsunami2} to integrate binary-single interactions. \textsc{tsunami}  employs the so-called algorithmic regularization chain scheme \citep[e.g.][]{ARChain, ARChain2, ARChain_improved}. This scheme is a combination of three numerical tecniques: Bulirsch-Stoer extrapolation \citep{bulirschStoer1964}, logarithmic Hamiltionian regularization \citep{ARChain2, pretoTremaine1999}, and chain-coordinate system \citep{aarseth1974, zare1974, heggie1974}. 

\textsc{tsunami}  also include velocity-dependent corrections to the Newtonian equations of motion, including post-Newtonian (PN) terms \citep[e.g.][]{PN_fack, PN_Blanchet} of order 1PN, 2PN and 2.5PN, and tidal forces \citep{press1977, 1981A&A....99..126H, tidesSamsing}.

In the next section we describe in detail the model for the dynamical tide that is included in \textsc{tsunami} .

\subsection{Dynamical tide model}
\textsc{tsunami}  uses the drag force model described in \cite{tidesSamsing}. The drag force is given by 
\begin{equation}
    \label{eq:tidalForce}
    \boldsymbol{F}=-\mathscr{E} \frac{v}{r^{n}} \times \frac{\mathbf{v}}{v}
\end{equation}
where $n=4$. $\mathscr{E}$ is a normalization factor that can be estimated with
\begin{equation}
    \label{eq:tidalCoeff}
    \mathscr{E}=\Delta E \times \frac{1}{2} \frac{\left[a\left(1-e^{2}\right)\right]^{n-1 / 2}}{(G M)^{1 / 2} \mathscr{I}(e, n)}
\end{equation}
$\mathscr{I}(e,n)$ is the solution to the integral 
\begin{equation}
    \label{eq:tidesIntegral}
    \mathscr{I}(e, n)=\int_{-\theta_{0}}^{+\theta_{0}} \frac{(1+e \cos \theta)-\left(1-e^{2}\right) / 2}{(1+e \cos \theta)^{2-n}} \mathrm{~d} \theta
\end{equation}
For any value on $n$, $\mathscr{I}$ can be written in a closed form. For $n = 4$, $\mathscr{I}$ evaluates to 
\begin{equation}
    \label{eq:tidesClosedForm}
    \mathscr{I}(e, n=4)= \frac{\pi}{2}\left(2+7 e^{2}+e^{4}\right),
\end{equation}
where $\theta_0$ is assumed to be $\pi$.

To obtain $\Delta E$ we consider a star with mass m$_1$ and radius r$_1$ on a parabolic orbit around a star with mass m$_2$, the energy loss due to tidal effects can be written as  \citep{tidalEnergyLoss}
\begin{equation}
\label{eq:tidalEnergyLoss}
    \Delta E = \frac{Gm_1^2}{r_1}\left(\frac{m_2}{m_1}\right)^2 \left(\frac{r_1}{r_p}\right)^6T_2(\eta)
\end{equation}
where $r_p$ is the pericenter distance of the orbit and $T_2$ is the tidal coupling constant of order 2. $T_l$, where $l>2$, is much smaller than $T_2$, to the point where $T_3$ is one order of magnitude less than $T_2$, and for higher $l$ it's even smaller. Therefore they are excluded and only $T_2$ is included in the code. For a polytropic star, the tidal coupling constant can be found by a polynomial fit \citep{tidalCouplingConstant}
\begin{equation}
\label{eq:tidalCouplingConstant}
    \log_{10} T_2(\eta) = \mathcal{A} + \mathcal{B}x + \mathcal{C}x^2 + \mathcal{D}x^3 + \mathcal{E}x^4 + \mathcal{F}x^5
\end{equation}
where $x = \log_{10} \eta$ and the coefficients, $\mathcal{A}, \mathcal{B}$, etc, can be found in table 1 of \citet{tidalCouplingConstant}. $\eta$ is calculated from \citep{tidalEnergyLoss}
\begin{equation}
    \eta = \left(\frac{m_1}{m_1 + m_2}\right)^{1/2}\left(\frac{r_p}{r_1}\right)^{3/2}
\end{equation}
These equations were extended to hyperbolic and eccentric orbits by \citet{tidalExtensionOtherOrbits}, where they introduced a new parameter $\zeta$, defined by
\begin{equation}
    \zeta = \eta \left(\frac{2}{1+e} \right)^{\alpha(\eta)/2}
\end{equation}
where $\alpha$ is given by
\begin{equation}
    \alpha = 1 + \frac{1}{2}\left|\frac{\eta - 2}{2}\right|^{3/2}
\end{equation}
This $\zeta$ is then used instead of $\eta$ in equations \ref{eq:tidalEnergyLoss} and \ref{eq:tidalCouplingConstant}.

The drag in force in \textsc{tsunami} is implemented in the following way, given two objects, $1$ and $2$, where tides on $1$ are excited by body $2$:

\begin{enumerate}[i.]
    \item Calculate $\mathscr{I}$ analytically with eq.\ref{eq:tidesClosedForm}.
    \item Calculate the semi-major axis and eccentricity of the orbit, assuming that the orbit is a Keplerian orbit.
    \item Use these values to calculate the energy loss using eq.\ref{eq:tidalEnergyLoss}.
    \item Estimate the drag force coefficient $\mathscr{E}$ using eq.\ref{eq:tidalCoeff}.
    \item Calculate the drag force vector between the two objects, $F_{1,2}$, using eq.\ref{eq:tidalForce}.
    \item The acceleration due to tidal energy losses on object 1 caused by object 2 is then calculated as $\mathbf{a}_{1,2} = \mathbf{F}_{1,2}/m_1$
    
\end{enumerate}
This process is repeated for all objects in the system and the resulting acceleration is the corresponding sum of the vectors. This is done at every time-step for all non-compact objects.

\subsection{Gravitational wave inspiral times}
While \textsc{tsunami}  includes PN corrections, it is not numerically convenient to directly integrate the coalescence of a isolated binary. For this reason, when a BH-BH binary survives an encounter, we estimate its coalescence time analytically.
An isolated BH binary will eventually merge due to energy losses through gravitational wave radiation emission. \citet{gravRadMergerTime} presents a way of approximating this merger time: 
\begin{equation}
\label{eq:GWMergers}
    t_{GW} = \frac{5c^5}{304G^3}\frac{a^4}{m_1 m_2 (m_1 + m_2)}f(e)
\end{equation} 
where c is the speed of light, a is the semi-major axis, e is the eccentricity, G is the gravitation constant and m$_1$ and m$_2$ are the masses of the two binary components. $f(e)$ is a factor which takes into account the binary eccentricity. As in \citet{2021arXiv211106388T}, we evaluate this factor numerically as:
\begin{equation}
    f(e) = \frac{(1-e^2)^4}{e^{\frac{48}{19}}(e^2 + \frac{304}{121})^{\frac{3480}{2299}}} \int_0^e \frac{x^{\frac{29}{19}}(1+\frac{121}{304}x^2)^{\frac{1181}{2299}}}{(1-x^2)^{3/2}} \;\mathrm{d}x
\end{equation}
 Through interactions, such as a flyby, with other objects, the semi-major axis and eccentricity can change. By comparing the merger time before and after the interaction we can determine if, on a statistical scale, interactions with other objects increase or decrease this merger time. This is relevant since the code itself may not classify the interaction as a merger but the resulting binary may merge shortly after the interaction if the semi-major axis was reduced or the eccentricity increased as a result of the third object interacting with the binary.

\begin{table*}
    \centering
    \caption{Table of possible outcomes for binary-single interactions involving two BHs and a star. Square brackets ("[ ]") indicate a bound system and a colon (":") indicates a merger. We make no distinction between a resonant or non-resonant flyby interaction so both non-resonant flybys and resonant flybys are counted as flybys. We also only report mergers as mergers, we do not show in this or future tables if the remaining two objects are bound or not.}
    \begin{center}
    \begin{tabular}{lcc}
    \toprule
        Outcome & BH$_1$ [BH$_2$ S] & S [BH$_1$ BH$_2$] \\
        \midrule
        Flyby & BH$_1$ [BH$_2$ S] & S [BH$_2$ BH$_1$] \\
        Exchange & S [BH$_1$ BH$_2$] or BH$_2$ [BH$_1$ S] &  BH$_2$ [BH$_1$ S] or BH$_1$ [BH$_2$ S] \\
        BH-BH merger & \multicolumn{2}{c}{BH$_1$:BH$_2$} \\
        BH-star merger & \multicolumn{2}{c}{BH$_1$:S or BH$_2$:S} \\
        Ionisation & \multicolumn{2}{c}{BH$_1$ BH$_2$ S} \\
        Stable triple & \multicolumn{2}{c}{$[[$BH$_1$ BH$_2$] S] or $[[$BH$_{1,2}$ S] BH$_{1,2}$]} \\
        \bottomrule
    \end{tabular}
    \end{center}
    \label{tab:outcomesTable}
\end{table*}

\subsection{Triple stability criterion}
\label{sec:tripleStability}
In order to classify a system as stable, \textsc{tsunami}  uses the approximate analytical criterion for dynamical stability from \citet{tidalExtensionOtherOrbits}. For a triple system with the objects 1, 2 and 3 we have an inner binary with $a_{\rm in}$ and $e_{\rm in}$ as well as a outer binary with $a_{\rm out}$ and $e_{\rm out}$. The mass ratio for the outer binary is $q_{out} = \frac{m_3}{(m_1 + m_2)}$ and we write the periastron separation as $R_{\rm p}^{\rm out}$. \citet{tidalExtensionOtherOrbits} gives the expression
\begin{equation}
\label{eq:tripleStability}
    \frac{R_{\text{p}}^{\text{out}}}{a_{\text{in}}} = C\left[(1+q_{\text{out}})\frac{1+e_{\text{out}}}{(1-e_{\text{out}})^{1/2}} \right]^{2/5} \equiv \frac{R_{\text{p}}^{\text{crit}}}{a_{\text{in}}}
\end{equation}
where $C=2.8$ is given empirically. A triple configuration will be considered stable if R$_p^{out}$ > R$_p^{crit}$ and the formula holds for $q \leq 5$.

\subsection{Initialisation \& outcomes of three-body encounters}

All interactions carried out in this work are done in 3D. \textsc{fewbody} is used to obtain the input data at $t=0$ which is then provided to \textsc{tsunami} in order to integrate the interactions. To generate the initial 3D positions and velocities coordinates at t=0: we provided fewbody with the masses and radii of the three stars involved in the interaction, the semi-major axis and eccentricity of the binary, the impact parameter and relative velocity of the interacting stars at infinity. The orbital elements of the interaction (inclination, longitude of the ascending node, argument of periastron and time of periastron passage of the binary) were randomly selected by \textsc{fewbody} as described in Section 3.1 of \cite{Fewbody_paper}. \textsc{fewbody} uses a seed to randomise these angles and for each interaction we select a seed from a large uniform distribution to ensure that the initial setup is unique to each interaction. we present each data set we mention how many seeds are used for each interaction. 

The MOnte Carlo Cluster simulAtor (MOCCA) \citep{MOCCA1, MOCCA2} is a one of the most advanced codes used to simulated real size star clusters such as globular clusters. We use several different globular cluster simulations\footnote{carried out as part of the \textsc{MOCCA}-Survey Database I \citep{askar2017}} to extract binary-single interactions involving two BHs and a single main-sequence star where the star is either as the initial single or is initially a binary component. We will discuss these sets in more detail in section \ref{sec:moccaSetup}.

\begin{figure*}
\resizebox{2.1\columnwidth}{!}{%
\tikzset{every picture/.style={line width=0.75pt}} 

\begin{tikzpicture}[x=0.75pt,y=0.75pt,yscale=-1,xscale=1]
\tikzstyle{every node}=[font=\Huge]

\draw  [fill={rgb, 255:red, 7; green, 0; blue, 0 }  ,fill opacity=1 ] (872,114) .. controls (872,100.19) and (883.19,89) .. (897,89) .. controls (910.81,89) and (922,100.19) .. (922,114) .. controls (922,127.81) and (910.81,139) .. (897,139) .. controls (883.19,139) and (872,127.81) .. (872,114) -- cycle ;
\draw  [fill={rgb, 255:red, 7; green, 0; blue, 0 }  ,fill opacity=1 ] (872,44) .. controls (872,30.19) and (883.19,19) .. (897,19) .. controls (910.81,19) and (922,30.19) .. (922,44) .. controls (922,57.81) and (910.81,69) .. (897,69) .. controls (883.19,69) and (872,57.81) .. (872,44) -- cycle ;
\draw  [fill={rgb, 255:red, 255; green, 255; blue, 255 }  ,fill opacity=1 ] (833,73.01) .. controls (833,52.57) and (847.42,36) .. (865.2,36) -- (865.2,45.66) .. controls (847.42,45.66) and (833,62.23) .. (833,82.67) ;\draw  [fill={rgb, 255:red, 255; green, 255; blue, 255 }  ,fill opacity=1 ] (833,82.67) .. controls (833,97.85) and (840.95,110.89) .. (852.32,116.6) -- (852.32,119.82) -- (865.2,114.85) -- (852.32,103.72) -- (852.32,106.94) .. controls (840.95,101.23) and (833,88.19) .. (833,73.01)(833,82.67) -- (833,73.01) ;
\draw  [fill={rgb, 255:red, 255; green, 255; blue, 255 }  ,fill opacity=1 ] (961.2,74.73) .. controls (961.2,94.95) and (946.78,111.34) .. (929,111.34) -- (929,121) .. controls (946.78,121) and (961.2,104.61) .. (961.2,84.39) ;\draw  [fill={rgb, 255:red, 255; green, 255; blue, 255 }  ,fill opacity=1 ] (961.2,84.39) .. controls (961.2,69.38) and (953.25,56.48) .. (941.88,50.83) -- (941.88,54.05) -- (929,42.95) -- (941.88,37.95) -- (941.88,41.17) .. controls (953.25,46.82) and (961.2,59.72) .. (961.2,74.73)(961.2,84.39) -- (961.2,74.73) ;
\draw   (1031.98,74.07) -- (1072.79,62.15) -- (1072.79,68.11) -- (1134,68.11) -- (1134,80.04) -- (1072.79,80.04) -- (1072.79,86) -- cycle ;
\draw   (790.06,196.89) -- (793.71,200.29) -- (844.01,146.16) -- (851.31,152.94) -- (801.01,207.07) -- (804.66,210.46) -- (763.83,239.76) -- cycle ;
\draw  [fill={rgb, 255:red, 7; green, 0; blue, 0 }  ,fill opacity=1 ] (690,361) .. controls (690,347.19) and (701.19,336) .. (715,336) .. controls (728.81,336) and (740,347.19) .. (740,361) .. controls (740,374.81) and (728.81,386) .. (715,386) .. controls (701.19,386) and (690,374.81) .. (690,361) -- cycle ;
\draw  [fill={rgb, 255:red, 7; green, 0; blue, 0 }  ,fill opacity=1 ] (690,291) .. controls (690,277.19) and (701.19,266) .. (715,266) .. controls (728.81,266) and (740,277.19) .. (740,291) .. controls (740,304.81) and (728.81,316) .. (715,316) .. controls (701.19,316) and (690,304.81) .. (690,291) -- cycle ;
\draw  [fill={rgb, 255:red, 255; green, 255; blue, 255 }  ,fill opacity=1 ] (651,320.01) .. controls (651,299.57) and (665.42,283) .. (683.2,283) -- (683.2,292.66) .. controls (665.42,292.66) and (651,309.23) .. (651,329.67) ;\draw  [fill={rgb, 255:red, 255; green, 255; blue, 255 }  ,fill opacity=1 ] (651,329.67) .. controls (651,344.85) and (658.95,357.89) .. (670.32,363.6) -- (670.32,366.82) -- (683.2,361.85) -- (670.32,350.72) -- (670.32,353.94) .. controls (658.95,348.23) and (651,335.19) .. (651,320.01)(651,329.67) -- (651,320.01) ;
\draw  [fill={rgb, 255:red, 255; green, 255; blue, 255 }  ,fill opacity=1 ] (779.2,321.73) .. controls (779.2,341.95) and (764.78,358.34) .. (747,358.34) -- (747,368) .. controls (764.78,368) and (779.2,351.61) .. (779.2,331.39) ;\draw  [fill={rgb, 255:red, 255; green, 255; blue, 255 }  ,fill opacity=1 ] (779.2,331.39) .. controls (779.2,316.38) and (771.25,303.48) .. (759.88,297.83) -- (759.88,301.05) -- (747,289.95) -- (759.88,284.95) -- (759.88,288.17) .. controls (771.25,293.82) and (779.2,306.72) .. (779.2,321.73)(779.2,331.39) -- (779.2,321.73) ;
\draw  [fill={rgb, 255:red, 7; green, 0; blue, 0 }  ,fill opacity=1 ] (1112,350) .. controls (1112,336.19) and (1123.19,325) .. (1137,325) .. controls (1150.81,325) and (1162,336.19) .. (1162,350) .. controls (1162,363.81) and (1150.81,375) .. (1137,375) .. controls (1123.19,375) and (1112,363.81) .. (1112,350) -- cycle ;
\draw  [fill={rgb, 255:red, 255; green, 255; blue, 255 }  ,fill opacity=1 ] (1073,309.01) .. controls (1073,288.57) and (1087.42,272) .. (1105.2,272) -- (1105.2,281.66) .. controls (1087.42,281.66) and (1073,298.23) .. (1073,318.67) ;\draw  [fill={rgb, 255:red, 255; green, 255; blue, 255 }  ,fill opacity=1 ] (1073,318.67) .. controls (1073,333.85) and (1080.95,346.89) .. (1092.32,352.6) -- (1092.32,355.82) -- (1105.2,350.85) -- (1092.32,339.72) -- (1092.32,342.94) .. controls (1080.95,337.23) and (1073,324.19) .. (1073,309.01)(1073,318.67) -- (1073,309.01) ;
\draw  [fill={rgb, 255:red, 255; green, 255; blue, 255 }  ,fill opacity=1 ] (1199,310.73) .. controls (1200.27,330.95) and (1186.88,347.34) .. (1169.09,347.34) -- (1169.7,357) .. controls (1187.48,357) and (1200.87,340.61) .. (1199.6,320.39) ;\draw  [fill={rgb, 255:red, 255; green, 255; blue, 255 }  ,fill opacity=1 ] (1199.6,320.39) .. controls (1198.66,305.38) and (1189.91,292.48) .. (1178.18,286.83) -- (1178.38,290.05) -- (1164.8,278.95) -- (1177.37,273.95) -- (1177.57,277.17) .. controls (1189.3,282.82) and (1198.06,295.72) .. (1199,310.73)(1199.6,320.39) -- (1199,310.73) ;
\draw   (1005.43,195.89) -- (1001.78,199.29) -- (951.48,145.16) -- (944.18,151.94) -- (994.48,206.07) -- (990.83,209.46) -- (1031.66,238.76) -- cycle ;
\draw   (526.88,13.5) -- (698.5,13.5) -- (698.5,165.5) -- (526.88,165.5) -- cycle ;
\draw  [fill={rgb, 255:red, 7; green, 0; blue, 0 }  ,fill opacity=1 ] (536,46) .. controls (536,32.19) and (547.19,21) .. (561,21) .. controls (574.81,21) and (586,32.19) .. (586,46) .. controls (586,59.81) and (574.81,71) .. (561,71) .. controls (547.19,71) and (536,59.81) .. (536,46) -- cycle ;
\draw  [fill={rgb, 255:red, 7; green, 0; blue, 0 }  ,fill opacity=1 ] (271.23,114) .. controls (271.23,100.19) and (282.43,89) .. (296.23,89) .. controls (310.04,89) and (321.23,100.19) .. (321.23,114) .. controls (321.23,127.81) and (310.04,139) .. (296.23,139) .. controls (282.43,139) and (271.23,127.81) .. (271.23,114) -- cycle ;
\draw  [fill={rgb, 255:red, 7; green, 0; blue, 0 }  ,fill opacity=1 ] (70.22,73) .. controls (70.22,59.19) and (59.02,48) .. (45.22,48) .. controls (31.41,48) and (20.22,59.19) .. (20.22,73) .. controls (20.22,86.81) and (31.41,98) .. (45.22,98) .. controls (59.02,98) and (70.22,86.81) .. (70.22,73) -- cycle ;
\draw  [fill={rgb, 255:red, 255; green, 255; blue, 255 }  ,fill opacity=1 ] (232.23,73.01) .. controls (232.23,52.57) and (246.65,36) .. (264.43,36) -- (264.43,45.66) .. controls (246.65,45.66) and (232.23,62.23) .. (232.23,82.67) ;\draw  [fill={rgb, 255:red, 255; green, 255; blue, 255 }  ,fill opacity=1 ] (232.23,82.67) .. controls (232.23,97.85) and (240.18,110.89) .. (251.55,116.6) -- (251.55,119.82) -- (264.43,114.85) -- (251.55,103.72) -- (251.55,106.94) .. controls (240.18,101.23) and (232.23,88.19) .. (232.23,73.01)(232.23,82.67) -- (232.23,73.01) ;
\draw  [fill={rgb, 255:red, 255; green, 255; blue, 255 }  ,fill opacity=1 ] (360.43,74.73) .. controls (360.43,94.95) and (346.02,111.34) .. (328.23,111.34) -- (328.23,121) .. controls (346.02,121) and (360.43,104.61) .. (360.43,84.39) ;\draw  [fill={rgb, 255:red, 255; green, 255; blue, 255 }  ,fill opacity=1 ] (360.43,84.39) .. controls (360.43,69.38) and (352.49,56.48) .. (341.11,50.83) -- (341.11,54.05) -- (328.23,42.95) -- (341.11,37.95) -- (341.11,41.17) .. controls (352.49,46.82) and (360.43,59.72) .. (360.43,74.73)(360.43,84.39) -- (360.43,74.73) ;
\draw   (194.23,74.07) -- (153.43,62.15) -- (153.43,68.11) -- (92.22,68.11) -- (92.22,80.04) -- (153.43,80.04) -- (153.43,86) -- cycle ;
\draw   (189.3,196.89) -- (192.95,200.29) -- (243.24,146.16) -- (250.54,152.94) -- (200.24,207.07) -- (203.89,210.46) -- (163.06,239.76) -- cycle ;
\draw  [fill={rgb, 255:red, 7; green, 0; blue, 0 }  ,fill opacity=1 ] (89.23,361) .. controls (89.23,347.19) and (100.43,336) .. (114.23,336) .. controls (128.04,336) and (139.23,347.19) .. (139.23,361) .. controls (139.23,374.81) and (128.04,386) .. (114.23,386) .. controls (100.43,386) and (89.23,374.81) .. (89.23,361) -- cycle ;
\draw  [fill={rgb, 255:red, 7; green, 0; blue, 0 }  ,fill opacity=1 ] (89.23,291) .. controls (89.23,277.19) and (100.43,266) .. (114.23,266) .. controls (128.04,266) and (139.23,277.19) .. (139.23,291) .. controls (139.23,304.81) and (128.04,316) .. (114.23,316) .. controls (100.43,316) and (89.23,304.81) .. (89.23,291) -- cycle ;
\draw  [fill={rgb, 255:red, 255; green, 255; blue, 255 }  ,fill opacity=1 ] (50.23,320.01) .. controls (50.23,299.57) and (64.65,283) .. (82.43,283) -- (82.43,292.66) .. controls (64.65,292.66) and (50.23,309.23) .. (50.23,329.67) ;\draw  [fill={rgb, 255:red, 255; green, 255; blue, 255 }  ,fill opacity=1 ] (50.23,329.67) .. controls (50.23,344.85) and (58.18,357.89) .. (69.55,363.6) -- (69.55,366.82) -- (82.43,361.85) -- (69.55,350.72) -- (69.55,353.94) .. controls (58.18,348.23) and (50.23,335.19) .. (50.23,320.01)(50.23,329.67) -- (50.23,320.01) ;
\draw  [fill={rgb, 255:red, 255; green, 255; blue, 255 }  ,fill opacity=1 ] (178.43,321.73) .. controls (178.43,341.95) and (164.02,358.34) .. (146.23,358.34) -- (146.23,368) .. controls (164.02,368) and (178.43,351.61) .. (178.43,331.39) ;\draw  [fill={rgb, 255:red, 255; green, 255; blue, 255 }  ,fill opacity=1 ] (178.43,331.39) .. controls (178.43,316.38) and (170.49,303.48) .. (159.11,297.83) -- (159.11,301.05) -- (146.23,289.95) -- (159.11,284.95) -- (159.11,288.17) .. controls (170.49,293.82) and (178.43,306.72) .. (178.43,321.73)(178.43,331.39) -- (178.43,321.73) ;
\draw  [fill={rgb, 255:red, 7; green, 0; blue, 0 }  ,fill opacity=1 ] (423.23,361) .. controls (423.23,347.19) and (434.43,336) .. (448.23,336) .. controls (462.04,336) and (473.23,347.19) .. (473.23,361) .. controls (473.23,374.81) and (462.04,386) .. (448.23,386) .. controls (434.43,386) and (423.23,374.81) .. (423.23,361) -- cycle ;
\draw  [fill={rgb, 255:red, 255; green, 255; blue, 255 }  ,fill opacity=1 ] (384.23,320.01) .. controls (384.23,299.57) and (398.65,283) .. (416.43,283) -- (416.43,292.66) .. controls (398.65,292.66) and (384.23,309.23) .. (384.23,329.67) ;\draw  [fill={rgb, 255:red, 255; green, 255; blue, 255 }  ,fill opacity=1 ] (384.23,329.67) .. controls (384.23,344.85) and (392.18,357.89) .. (403.55,363.6) -- (403.55,366.82) -- (416.43,361.85) -- (403.55,350.72) -- (403.55,353.94) .. controls (392.18,348.23) and (384.23,335.19) .. (384.23,320.01)(384.23,329.67) -- (384.23,320.01) ;
\draw  [fill={rgb, 255:red, 255; green, 255; blue, 255 }  ,fill opacity=1 ] (512.43,321.73) .. controls (512.43,341.95) and (498.02,358.34) .. (480.23,358.34) -- (480.23,368) .. controls (498.02,368) and (512.43,351.61) .. (512.43,331.39) ;\draw  [fill={rgb, 255:red, 255; green, 255; blue, 255 }  ,fill opacity=1 ] (512.43,331.39) .. controls (512.43,316.38) and (504.49,303.48) .. (493.11,297.83) -- (493.11,301.05) -- (480.23,289.95) -- (493.11,284.95) -- (493.11,288.17) .. controls (504.49,293.82) and (512.43,306.72) .. (512.43,321.73)(512.43,331.39) -- (512.43,321.73) ;
\draw  [fill={rgb, 255:red, 7; green, 0; blue, 0 }  ,fill opacity=1 ] (113.43,501.66) .. controls (113.43,487.85) and (124.63,476.66) .. (138.43,476.66) .. controls (152.24,476.66) and (163.43,487.85) .. (163.43,501.66) .. controls (163.43,515.47) and (152.24,526.66) .. (138.43,526.66) .. controls (124.63,526.66) and (113.43,515.47) .. (113.43,501.66) -- cycle ;
\draw  [fill={rgb, 255:red, 255; green, 255; blue, 255 }  ,fill opacity=1 ] (81.23,529.01) .. controls (81.23,508.57) and (95.65,492) .. (113.43,492) -- (113.43,501.66) .. controls (95.65,501.66) and (81.23,518.23) .. (81.23,538.67) ;\draw  [fill={rgb, 255:red, 255; green, 255; blue, 255 }  ,fill opacity=1 ] (81.23,538.67) .. controls (81.23,553.85) and (89.18,566.89) .. (100.55,572.6) -- (100.55,575.82) -- (113.43,570.85) -- (100.55,559.72) -- (100.55,562.94) .. controls (89.18,557.23) and (81.23,544.19) .. (81.23,529.01)(81.23,538.67) -- (81.23,529.01) ;
\draw  [fill={rgb, 255:red, 255; green, 255; blue, 255 }  ,fill opacity=1 ] (209.43,530.73) .. controls (209.43,550.95) and (195.02,567.34) .. (177.23,567.34) -- (177.23,577) .. controls (195.02,577) and (209.43,560.61) .. (209.43,540.39) ;\draw  [fill={rgb, 255:red, 255; green, 255; blue, 255 }  ,fill opacity=1 ] (209.43,540.39) .. controls (209.43,525.38) and (201.49,512.48) .. (190.11,506.83) -- (190.11,510.05) -- (177.23,498.95) -- (190.11,493.95) -- (190.11,497.17) .. controls (201.49,502.82) and (209.43,515.72) .. (209.43,530.73)(209.43,540.39) -- (209.43,530.73) ;
\draw   (283.37,323.26) -- (289.99,323.35) -- (292.52,147.04) -- (305.75,147.23) -- (303.23,323.54) -- (309.85,323.64) -- (294.93,440.99) -- cycle ;
\draw   (404.66,195.89) -- (401.01,199.29) -- (350.71,145.16) -- (343.41,151.94) -- (393.71,206.07) -- (390.06,209.46) -- (430.89,238.76) -- cycle ;
\draw  [fill={rgb, 255:red, 7; green, 0; blue, 0 }  ,fill opacity=1 ] (133.43,502.66) .. controls (133.43,488.85) and (144.63,477.66) .. (158.43,477.66) .. controls (172.24,477.66) and (183.43,488.85) .. (183.43,502.66) .. controls (183.43,516.47) and (172.24,527.66) .. (158.43,527.66) .. controls (144.63,527.66) and (133.43,516.47) .. (133.43,502.66) -- cycle ;
\draw  [fill={rgb, 255:red, 252; green, 233; blue, 0 }  ,fill opacity=0.98 ] (298.23,2.5) -- (308.52,23.04) -- (331.52,26.34) -- (314.88,42.33) -- (318.81,64.91) -- (298.23,54.25) -- (277.66,64.91) -- (281.59,42.33) -- (264.95,26.34) -- (287.95,23.04) -- cycle ;
\draw  [fill={rgb, 255:red, 7; green, 0; blue, 0 }  ,fill opacity=1 ] (269.23,573) .. controls (269.23,559.19) and (280.43,548) .. (294.23,548) .. controls (308.04,548) and (319.23,559.19) .. (319.23,573) .. controls (319.23,586.81) and (308.04,598) .. (294.23,598) .. controls (280.43,598) and (269.23,586.81) .. (269.23,573) -- cycle ;
\draw  [fill={rgb, 255:red, 7; green, 0; blue, 0 }  ,fill opacity=1 ] (262.43,504.66) .. controls (262.43,490.85) and (273.63,479.66) .. (287.43,479.66) .. controls (301.24,479.66) and (312.43,490.85) .. (312.43,504.66) .. controls (312.43,518.47) and (301.24,529.66) .. (287.43,529.66) .. controls (273.63,529.66) and (262.43,518.47) .. (262.43,504.66) -- cycle ;
\draw  [fill={rgb, 255:red, 255; green, 255; blue, 255 }  ,fill opacity=1 ] (230.23,532.01) .. controls (230.23,511.57) and (244.65,495) .. (262.43,495) -- (262.43,504.66) .. controls (244.65,504.66) and (230.23,521.23) .. (230.23,541.67) ;\draw  [fill={rgb, 255:red, 255; green, 255; blue, 255 }  ,fill opacity=1 ] (230.23,541.67) .. controls (230.23,556.85) and (238.18,569.89) .. (249.55,575.6) -- (249.55,578.82) -- (262.43,573.85) -- (249.55,562.72) -- (249.55,565.94) .. controls (238.18,560.23) and (230.23,547.19) .. (230.23,532.01)(230.23,541.67) -- (230.23,532.01) ;
\draw  [fill={rgb, 255:red, 7; green, 0; blue, 0 }  ,fill opacity=1 ] (426.23,576) .. controls (426.23,562.19) and (437.43,551) .. (451.23,551) .. controls (465.04,551) and (476.23,562.19) .. (476.23,576) .. controls (476.23,589.81) and (465.04,601) .. (451.23,601) .. controls (437.43,601) and (426.23,589.81) .. (426.23,576) -- cycle ;
\draw  [fill={rgb, 255:red, 7; green, 0; blue, 0 }  ,fill opacity=1 ] (419.43,507.66) .. controls (419.43,493.85) and (430.63,482.66) .. (444.43,482.66) .. controls (458.24,482.66) and (469.43,493.85) .. (469.43,507.66) .. controls (469.43,521.47) and (458.24,532.66) .. (444.43,532.66) .. controls (430.63,532.66) and (419.43,521.47) .. (419.43,507.66) -- cycle ;
\draw  [fill={rgb, 255:red, 255; green, 255; blue, 255 }  ,fill opacity=1 ] (387.23,535.01) .. controls (387.23,514.57) and (401.65,498) .. (419.43,498) -- (419.43,507.66) .. controls (401.65,507.66) and (387.23,524.23) .. (387.23,544.67) ;\draw  [fill={rgb, 255:red, 255; green, 255; blue, 255 }  ,fill opacity=1 ] (387.23,544.67) .. controls (387.23,559.85) and (395.18,572.89) .. (406.55,578.6) -- (406.55,581.82) -- (419.43,576.85) -- (406.55,565.72) -- (406.55,568.94) .. controls (395.18,563.23) and (387.23,550.19) .. (387.23,535.01)(387.23,544.67) -- (387.23,535.01) ;
\draw  [fill={rgb, 255:red, 252; green, 233; blue, 0 }  ,fill opacity=0.98 ] (1170.23,34.5) -- (1180.52,55.04) -- (1203.52,58.34) -- (1186.88,74.33) -- (1190.81,96.91) -- (1170.23,86.25) -- (1149.66,96.91) -- (1153.59,74.33) -- (1136.95,58.34) -- (1159.95,55.04) -- cycle ;
\draw  [fill={rgb, 255:red, 252; green, 233; blue, 0 }  ,fill opacity=0.98 ] (448.23,251.5) -- (458.52,272.04) -- (481.52,275.34) -- (464.88,291.33) -- (468.81,313.91) -- (448.23,303.25) -- (427.66,313.91) -- (431.59,291.33) -- (414.95,275.34) -- (437.95,272.04) -- cycle ;
\draw  [fill={rgb, 255:red, 252; green, 233; blue, 0 }  ,fill opacity=0.98 ] (146.23,532.5) -- (156.52,553.04) -- (179.52,556.34) -- (162.88,572.33) -- (166.81,594.91) -- (146.23,584.25) -- (125.66,594.91) -- (129.59,572.33) -- (112.95,556.34) -- (135.95,553.04) -- cycle ;
\draw  [fill={rgb, 255:red, 252; green, 233; blue, 0 }  ,fill opacity=0.98 ] (308.01,467.25) -- (318.29,487.79) -- (341.29,491.09) -- (324.65,507.08) -- (328.58,529.66) -- (308.01,519) -- (287.43,529.66) -- (291.36,507.08) -- (274.72,491.09) -- (297.72,487.79) -- cycle ;
\draw  [fill={rgb, 255:red, 252; green, 233; blue, 0 }  ,fill opacity=0.98 ] (465.01,470.25) -- (475.29,490.79) -- (498.29,494.09) -- (481.65,510.08) -- (485.58,532.66) -- (465.01,522) -- (444.43,532.66) -- (448.36,510.08) -- (431.72,494.09) -- (454.72,490.79) -- cycle ;
\draw  [fill={rgb, 255:red, 255; green, 255; blue, 255 }  ,fill opacity=1 ] (358.43,533.73) .. controls (358.43,553.95) and (344.02,570.34) .. (326.23,570.34) -- (326.23,580) .. controls (344.02,580) and (358.43,563.61) .. (358.43,543.39) ;\draw  [fill={rgb, 255:red, 255; green, 255; blue, 255 }  ,fill opacity=1 ] (358.43,543.39) .. controls (358.43,528.38) and (350.49,515.48) .. (339.11,509.83) -- (339.11,513.05) -- (326.23,501.95) -- (339.11,496.95) -- (339.11,500.17) .. controls (350.49,505.82) and (358.43,518.72) .. (358.43,533.73)(358.43,543.39) -- (358.43,533.73) ;
\draw  [fill={rgb, 255:red, 255; green, 255; blue, 255 }  ,fill opacity=1 ] (515.43,536.73) .. controls (515.43,556.95) and (501.02,573.34) .. (483.23,573.34) -- (483.23,583) .. controls (501.02,583) and (515.43,566.61) .. (515.43,546.39) ;\draw  [fill={rgb, 255:red, 255; green, 255; blue, 255 }  ,fill opacity=1 ] (515.43,546.39) .. controls (515.43,531.38) and (507.49,518.48) .. (496.11,512.83) -- (496.11,516.05) -- (483.23,504.95) -- (496.11,499.95) -- (496.11,503.17) .. controls (507.49,508.82) and (515.43,521.72) .. (515.43,536.73)(515.43,546.39) -- (515.43,536.73) ;
\draw  [fill={rgb, 255:red, 252; green, 233; blue, 0 }  ,fill opacity=0.98 ] (562.23,97.5) -- (572.52,118.04) -- (595.52,121.34) -- (578.88,137.33) -- (582.81,159.91) -- (562.23,149.25) -- (541.66,159.91) -- (545.59,137.33) -- (528.95,121.34) -- (551.95,118.04) -- cycle ;
\draw  [fill={rgb, 255:red, 252; green, 233; blue, 0 }  ,fill opacity=0.98 ] (1135,240.5) -- (1145.29,261.04) -- (1168.29,264.34) -- (1151.64,280.33) -- (1155.57,302.91) -- (1135,292.25) -- (1114.43,302.91) -- (1118.36,280.33) -- (1101.71,264.34) -- (1124.71,261.04) -- cycle ;
\draw  [fill={rgb, 255:red, 7; green, 0; blue, 0 }  ,fill opacity=1 ] (972.8,269) .. controls (972.8,255.19) and (983.99,244) .. (997.8,244) .. controls (1011.61,244) and (1022.8,255.19) .. (1022.8,269) .. controls (1022.8,282.81) and (1011.61,294) .. (997.8,294) .. controls (983.99,294) and (972.8,282.81) .. (972.8,269) -- cycle ;
\draw  [fill={rgb, 255:red, 255; green, 255; blue, 255 }  ,fill opacity=1 ] (931.8,304.01) .. controls (931.8,283.57) and (946.22,267) .. (964,267) -- (964,276.66) .. controls (946.22,276.66) and (931.8,293.23) .. (931.8,313.67) ;\draw  [fill={rgb, 255:red, 255; green, 255; blue, 255 }  ,fill opacity=1 ] (931.8,313.67) .. controls (931.8,328.85) and (939.75,341.89) .. (951.12,347.6) -- (951.12,350.82) -- (964,345.85) -- (951.12,334.72) -- (951.12,337.94) .. controls (939.75,332.23) and (931.8,319.19) .. (931.8,304.01)(931.8,313.67) -- (931.8,304.01) ;
\draw  [fill={rgb, 255:red, 255; green, 255; blue, 255 }  ,fill opacity=1 ] (1057.8,305.73) .. controls (1059.07,325.95) and (1045.68,342.34) .. (1027.89,342.34) -- (1028.5,352) .. controls (1046.28,352) and (1059.67,335.61) .. (1058.4,315.39) ;\draw  [fill={rgb, 255:red, 255; green, 255; blue, 255 }  ,fill opacity=1 ] (1058.4,315.39) .. controls (1057.46,300.38) and (1048.71,287.48) .. (1036.98,281.83) -- (1037.18,285.05) -- (1023.6,273.95) -- (1036.17,268.95) -- (1036.37,272.17) .. controls (1048.1,277.82) and (1056.86,290.72) .. (1057.8,305.73)(1058.4,315.39) -- (1057.8,305.73) ;
\draw  [fill={rgb, 255:red, 252; green, 233; blue, 0 }  ,fill opacity=0.98 ] (996.8,308.5) -- (1007.09,329.04) -- (1030.09,332.34) -- (1013.44,348.33) -- (1017.37,370.91) -- (996.8,360.25) -- (976.23,370.91) -- (980.16,348.33) -- (963.51,332.34) -- (986.51,329.04) -- cycle ;
\draw   (882.37,333.26) -- (888.99,333.35) -- (891.52,157.04) -- (904.75,157.23) -- (902.23,333.54) -- (908.85,333.64) -- (893.93,450.99) -- cycle ;
\draw  [fill={rgb, 255:red, 7; green, 0; blue, 0 }  ,fill opacity=1 ] (718.43,503.66) .. controls (718.43,489.85) and (729.63,478.66) .. (743.43,478.66) .. controls (757.24,478.66) and (768.43,489.85) .. (768.43,503.66) .. controls (768.43,517.47) and (757.24,528.66) .. (743.43,528.66) .. controls (729.63,528.66) and (718.43,517.47) .. (718.43,503.66) -- cycle ;
\draw  [fill={rgb, 255:red, 255; green, 255; blue, 255 }  ,fill opacity=1 ] (686.23,531.01) .. controls (686.23,510.57) and (700.65,494) .. (718.43,494) -- (718.43,503.66) .. controls (700.65,503.66) and (686.23,520.23) .. (686.23,540.67) ;\draw  [fill={rgb, 255:red, 255; green, 255; blue, 255 }  ,fill opacity=1 ] (686.23,540.67) .. controls (686.23,555.85) and (694.18,568.89) .. (705.55,574.6) -- (705.55,577.82) -- (718.43,572.85) -- (705.55,561.72) -- (705.55,564.94) .. controls (694.18,559.23) and (686.23,546.19) .. (686.23,531.01)(686.23,540.67) -- (686.23,531.01) ;
\draw  [fill={rgb, 255:red, 255; green, 255; blue, 255 }  ,fill opacity=1 ] (814.43,532.73) .. controls (814.43,552.95) and (800.02,569.34) .. (782.23,569.34) -- (782.23,579) .. controls (800.02,579) and (814.43,562.61) .. (814.43,542.39) ;\draw  [fill={rgb, 255:red, 255; green, 255; blue, 255 }  ,fill opacity=1 ] (814.43,542.39) .. controls (814.43,527.38) and (806.49,514.48) .. (795.11,508.83) -- (795.11,512.05) -- (782.23,500.95) -- (795.11,495.95) -- (795.11,499.17) .. controls (806.49,504.82) and (814.43,517.72) .. (814.43,532.73)(814.43,542.39) -- (814.43,532.73) ;
\draw  [fill={rgb, 255:red, 7; green, 0; blue, 0 }  ,fill opacity=1 ] (738.43,504.66) .. controls (738.43,490.85) and (749.63,479.66) .. (763.43,479.66) .. controls (777.24,479.66) and (788.43,490.85) .. (788.43,504.66) .. controls (788.43,518.47) and (777.24,529.66) .. (763.43,529.66) .. controls (749.63,529.66) and (738.43,518.47) .. (738.43,504.66) -- cycle ;
\draw  [fill={rgb, 255:red, 7; green, 0; blue, 0 }  ,fill opacity=1 ] (874.23,575) .. controls (874.23,561.19) and (885.43,550) .. (899.23,550) .. controls (913.04,550) and (924.23,561.19) .. (924.23,575) .. controls (924.23,588.81) and (913.04,600) .. (899.23,600) .. controls (885.43,600) and (874.23,588.81) .. (874.23,575) -- cycle ;
\draw  [fill={rgb, 255:red, 7; green, 0; blue, 0 }  ,fill opacity=1 ] (867.43,506.66) .. controls (867.43,492.85) and (878.63,481.66) .. (892.43,481.66) .. controls (906.24,481.66) and (917.43,492.85) .. (917.43,506.66) .. controls (917.43,520.47) and (906.24,531.66) .. (892.43,531.66) .. controls (878.63,531.66) and (867.43,520.47) .. (867.43,506.66) -- cycle ;
\draw  [fill={rgb, 255:red, 255; green, 255; blue, 255 }  ,fill opacity=1 ] (835.23,534.01) .. controls (835.23,513.57) and (849.65,497) .. (867.43,497) -- (867.43,506.66) .. controls (849.65,506.66) and (835.23,523.23) .. (835.23,543.67) ;\draw  [fill={rgb, 255:red, 255; green, 255; blue, 255 }  ,fill opacity=1 ] (835.23,543.67) .. controls (835.23,558.85) and (843.18,571.89) .. (854.55,577.6) -- (854.55,580.82) -- (867.43,575.85) -- (854.55,564.72) -- (854.55,567.94) .. controls (843.18,562.23) and (835.23,549.19) .. (835.23,534.01)(835.23,543.67) -- (835.23,534.01) ;
\draw  [fill={rgb, 255:red, 7; green, 0; blue, 0 }  ,fill opacity=1 ] (1031.23,578) .. controls (1031.23,564.19) and (1042.43,553) .. (1056.23,553) .. controls (1070.04,553) and (1081.23,564.19) .. (1081.23,578) .. controls (1081.23,591.81) and (1070.04,603) .. (1056.23,603) .. controls (1042.43,603) and (1031.23,591.81) .. (1031.23,578) -- cycle ;
\draw  [fill={rgb, 255:red, 7; green, 0; blue, 0 }  ,fill opacity=1 ] (1024.43,509.66) .. controls (1024.43,495.85) and (1035.63,484.66) .. (1049.43,484.66) .. controls (1063.24,484.66) and (1074.43,495.85) .. (1074.43,509.66) .. controls (1074.43,523.47) and (1063.24,534.66) .. (1049.43,534.66) .. controls (1035.63,534.66) and (1024.43,523.47) .. (1024.43,509.66) -- cycle ;
\draw  [fill={rgb, 255:red, 255; green, 255; blue, 255 }  ,fill opacity=1 ] (992.23,537.01) .. controls (992.23,516.57) and (1006.65,500) .. (1024.43,500) -- (1024.43,509.66) .. controls (1006.65,509.66) and (992.23,526.23) .. (992.23,546.67) ;\draw  [fill={rgb, 255:red, 255; green, 255; blue, 255 }  ,fill opacity=1 ] (992.23,546.67) .. controls (992.23,561.85) and (1000.18,574.89) .. (1011.55,580.6) -- (1011.55,583.82) -- (1024.43,578.85) -- (1011.55,567.72) -- (1011.55,570.94) .. controls (1000.18,565.23) and (992.23,552.19) .. (992.23,537.01)(992.23,546.67) -- (992.23,537.01) ;
\draw  [fill={rgb, 255:red, 252; green, 233; blue, 0 }  ,fill opacity=0.98 ] (751.23,534.5) -- (761.52,555.04) -- (784.52,558.34) -- (767.88,574.33) -- (771.81,596.91) -- (751.23,586.25) -- (730.66,596.91) -- (734.59,574.33) -- (717.95,558.34) -- (740.95,555.04) -- cycle ;
\draw  [fill={rgb, 255:red, 252; green, 233; blue, 0 }  ,fill opacity=0.98 ] (913.01,469.25) -- (923.29,489.79) -- (946.29,493.09) -- (929.65,509.08) -- (933.58,531.66) -- (913.01,521) -- (892.43,531.66) -- (896.36,509.08) -- (879.72,493.09) -- (902.72,489.79) -- cycle ;
\draw  [fill={rgb, 255:red, 252; green, 233; blue, 0 }  ,fill opacity=0.98 ] (1070.01,472.25) -- (1080.29,492.79) -- (1103.29,496.09) -- (1086.65,512.08) -- (1090.58,534.66) -- (1070.01,524) -- (1049.43,534.66) -- (1053.36,512.08) -- (1036.72,496.09) -- (1059.72,492.79) -- cycle ;
\draw  [fill={rgb, 255:red, 255; green, 255; blue, 255 }  ,fill opacity=1 ] (963.43,535.73) .. controls (963.43,555.95) and (949.02,572.34) .. (931.23,572.34) -- (931.23,582) .. controls (949.02,582) and (963.43,565.61) .. (963.43,545.39) ;\draw  [fill={rgb, 255:red, 255; green, 255; blue, 255 }  ,fill opacity=1 ] (963.43,545.39) .. controls (963.43,530.38) and (955.49,517.48) .. (944.11,511.83) -- (944.11,515.05) -- (931.23,503.95) -- (944.11,498.95) -- (944.11,502.17) .. controls (955.49,507.82) and (963.43,520.72) .. (963.43,535.73)(963.43,545.39) -- (963.43,535.73) ;
\draw  [fill={rgb, 255:red, 255; green, 255; blue, 255 }  ,fill opacity=1 ] (1120.43,538.73) .. controls (1120.43,558.95) and (1106.02,575.34) .. (1088.23,575.34) -- (1088.23,585) .. controls (1106.02,585) and (1120.43,568.61) .. (1120.43,548.39) ;\draw  [fill={rgb, 255:red, 255; green, 255; blue, 255 }  ,fill opacity=1 ] (1120.43,548.39) .. controls (1120.43,533.38) and (1112.49,520.48) .. (1101.11,514.83) -- (1101.11,518.05) -- (1088.23,506.95) -- (1101.11,501.95) -- (1101.11,505.17) .. controls (1112.49,510.82) and (1120.43,523.72) .. (1120.43,538.73)(1120.43,548.39) -- (1120.43,538.73) ;

\draw (595,35) node [anchor=north west][inner sep=0.75pt]   [align=left] {{\huge Black hole}};
\draw (612,120) node [anchor=north west][inner sep=0.75pt]   [align=left] {{\huge Star}};
\draw (404,387) node [anchor=north west][inner sep=0.75pt]   [align=left] {{\Huge Flyby}};
\draw (60,390) node [anchor=north west][inner sep=0.75pt]   [align=left] {{\Huge Exchange}};
\draw (675,392) node [anchor=north west][inner sep=0.75pt]   [align=left] {{\Huge Flyby}};
\draw (991,384) node [anchor=north west][inner sep=0.75pt]   [align=left] {{\Huge Exchange}};
\draw (36,60) node [anchor=north west][inner sep=0.75pt]  [color={rgb, 255:red, 255; green, 255; blue, 255 }  ,opacity=1 ] [align=left] {{\Huge 1}};
\draw (105,278) node [anchor=north west][inner sep=0.75pt]  [color={rgb, 255:red, 255; green, 255; blue, 255 }  ,opacity=1 ] [align=left] {{\Huge 1}};
\draw (287,103) node [anchor=north west][inner sep=0.75pt]  [color={rgb, 255:red, 255; green, 255; blue, 255 }  ,opacity=1 ] [align=left] {{\Huge 2}};
\draw (106,350) node [anchor=north west][inner sep=0.75pt]  [color={rgb, 255:red, 255; green, 255; blue, 255 }  ,opacity=1 ] [align=left] {{\Huge 2}};
\draw (440,350) node [anchor=north west][inner sep=0.75pt]  [color={rgb, 255:red, 255; green, 255; blue, 255 }  ,opacity=1 ] [align=left] {{\Huge 2}};
\draw (889,31) node [anchor=north west][inner sep=0.75pt]  [color={rgb, 255:red, 255; green, 255; blue, 255 }  ,opacity=1 ] [align=left] {{\Huge 1}};
\draw (707,279) node [anchor=north west][inner sep=0.75pt]  [color={rgb, 255:red, 255; green, 255; blue, 255 }  ,opacity=1 ] [align=left] {{\Huge 1}};
\draw (1128,339) node [anchor=north west][inner sep=0.75pt]  [color={rgb, 255:red, 255; green, 255; blue, 255 }  ,opacity=1 ] [align=left] {{\Huge 2}};
\draw (707,350) node [anchor=north west][inner sep=0.75pt]  [color={rgb, 255:red, 255; green, 255; blue, 255 }  ,opacity=1 ] [align=left] {{\Huge 2}};
\draw (889,102) node [anchor=north west][inner sep=0.75pt]  [color={rgb, 255:red, 255; green, 255; blue, 255 }  ,opacity=1 ] [align=left] {{\Huge 2}};
\draw (245,617) node [anchor=north west][inner sep=0.75pt]   [align=left] {{\Huge Mergers}};
\draw (286,561) node [anchor=north west][inner sep=0.75pt]  [color={rgb, 255:red, 255; green, 255; blue, 255 }  ,opacity=1 ] [align=left] {{\Huge 2}};
\draw (443,563) node [anchor=north west][inner sep=0.75pt]  [color={rgb, 255:red, 255; green, 255; blue, 255 }  ,opacity=1 ] [align=left] {{\Huge 1}};
\draw (122,488) node [anchor=north west][inner sep=0.75pt]  [color={rgb, 255:red, 255; green, 255; blue, 255 }  ,opacity=1 ] [align=left] {{\Huge 1+2}};
\draw (264,492) node [anchor=north west][inner sep=0.75pt]  [color={rgb, 255:red, 255; green, 255; blue, 255 }  ,opacity=1 ] [align=left] {{\Huge 1}};
\draw (422,492) node [anchor=north west][inner sep=0.75pt]  [color={rgb, 255:red, 255; green, 255; blue, 255 }  ,opacity=1 ] [align=left] {{\Huge 2}};
\draw (989,256) node [anchor=north west][inner sep=0.75pt]  [color={rgb, 255:red, 255; green, 255; blue, 255 }  ,opacity=1 ] [align=left] {{\Huge 1}};
\draw (850,619) node [anchor=north west][inner sep=0.75pt]   [align=left] {{\Huge Mergers}};
\draw (891,562) node [anchor=north west][inner sep=0.75pt]  [color={rgb, 255:red, 255; green, 255; blue, 255 }  ,opacity=1 ] [align=left] {{\Huge 2}};
\draw (1048,565) node [anchor=north west][inner sep=0.75pt]  [color={rgb, 255:red, 255; green, 255; blue, 255 }  ,opacity=1 ] [align=left] {{\Huge 1}};
\draw (727,489) node [anchor=north west][inner sep=0.75pt]  [color={rgb, 255:red, 255; green, 255; blue, 255 }  ,opacity=1 ] [align=left] {{\Huge 1+2}};
\draw (869,498) node [anchor=north west][inner sep=0.75pt]  [color={rgb, 255:red, 255; green, 255; blue, 255 }  ,opacity=1 ] [align=left] {{\Huge 1}};
\draw (1027,498) node [anchor=north west][inner sep=0.75pt]  [color={rgb, 255:red, 255; green, 255; blue, 255 }  ,opacity=1 ] [align=left] {{\Huge 2}};

\end{tikzpicture}
}
\caption{Schematic of possible outcomes in a binary-single interaction. The left side of the plot shows a BH-star binary interacting with a single BH and the right side shows a BH-BH binary interacting with a single star. Possible outcomes are an exchange, a flyby or a merger. Our results also show a small number of Ionisations where the end result is three single objects but this is not shown in the above schematic. Long lived stable triples can also be formed but are usually broken up quickly in clusters due to the high stellar density and relatively high encounter rate}
\label{fig:outcomesSchematics}
\end{figure*}
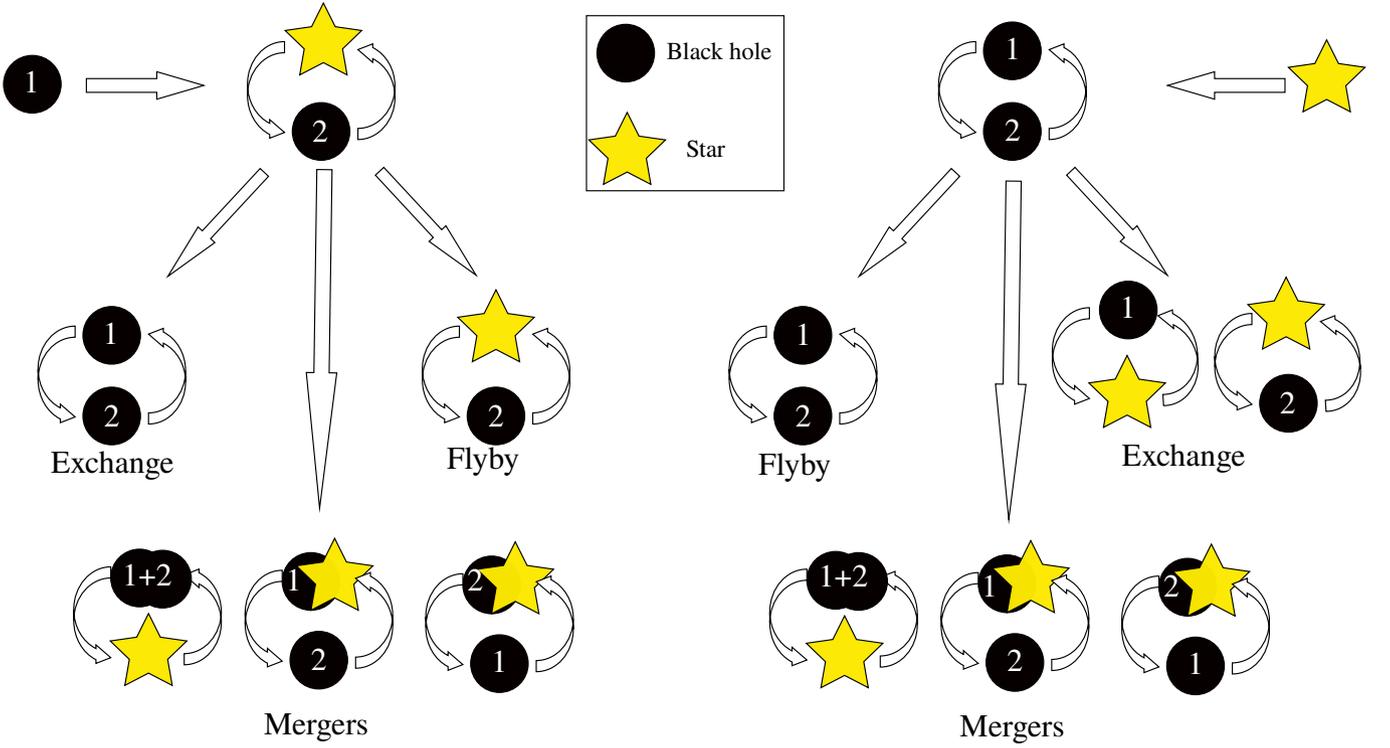

Figure \ref{fig:outcomesSchematics} shows a schematic of outcomes during a binary-single interaction with two BHs and a star. The schematic is split into two sides, on the left side, the initial binary contains a star and a BH and the single is a BH. On the right side, the initial binary is a BH binary and the single is a star. The vertical lines represent the different outcomes that can occur: flyby, exchange or a merger. There is also a possibility that the single breaks up the binary, an ionisation, and the result is 3 singles. Another outcome not shown in the schematic is a stable hierarchical triple where we have an inner binary with the third object bound to this binary. Both of these outcomes are  uncommon in our data set and are not included in the schematic. For all outcomes see table \ref{tab:outcomesTable}.

We use notations similar to what is used in \citet{Fewbody_paper} for bound systems and mergers. A bound system is represented by "[ ]", e.x. "BH$_1$ [BH$_2$ S]" represents a binary with a BH and a star with an incoming BH. Mergers are represented by a ":", for example, "BH$_1$:BH$_2$" represents a BH merger and "BH$_1$:S" represents a merger between the incoming BH and the star.

\subsection{Binary mergers due to tidal dissipation}
\label{sec:instantMergers}

With the dynamical tides treatment in \textsc{tsunami} , a few BH-Star binaries with very high initial eccentricities would merge in an inspiral caused by the tidal effects before the single BH gets close. Since these mergers cannot be considered to be three-body interactions we need to filter them out. We do this by considering the binary to be a single body and we can thus work with analytical formulas for two-body interactions. We apply three criteria in order to find these `instant' mergers:
\begin{enumerate}
    \item The merger occurs between the two initial binary components and one of the objects is a star.
    \item The merger time returned by \textsc{tsunami}  is less than half of the time to pericenter for the orbit between the binary and the single.
    \item The separation between the COM of the two merger objects and the third object is greater than 5 times the initial semi-major axis of the binary.
\end{enumerate}
    
Any interaction which fulfills these three criteria are considered to be a merger not influenced by the single and filtered out and not included in our results reported in this paper.

\subsection{Merger criterion}
    Mergers in \textsc{tsunami}  are classified using the ``sticky star'' approximation. When the separation between two stars are less than the sum of their radii, \textsc{tsunami}  stops the interaction and records that a merger occured during the interaction.

\section{Test setup}
\label{sec:testSetup}
\subsection{Initial properties}
In order to see the effect of the different flags in a more controlled set we create a test setup with initial properties as seen in table \ref{tab:testSetup_initalProperties}. All stars in this set are Sun-like stars while the BHs have a mass of 10 M$_{\odot}$.  The semi-major axis is binned from a minimum of 5 R$_{\odot}$ up to a maximum of 1 AU with a total of 8 uniformly distributed bins. For simplicity we set the eccentricity to 0 and V$_{\infty}$ to 5 km/s. We are mostly interested in strong encounters and need to set our impact parameter accordingly. We do this by considering gravitational focusing in the limit where the velocity at pericenter is much higher than the velocity at infinity \citep{samsingBMaxCitation}. This allows us to get a maximum value of the impact parameter using equation \ref{eq:bMax} where G is the gravitational constant, M$_{\mathrm{tot}}$ is the total mass of the three objects, $r_{peri}$ is the pericenter distance between the binary and the single and $V_{\infty}$ is the velocity at infinity. 
\begin{equation}
\label{eq:bMax}
    b_{\text{max}} = \sqrt{\frac{2G \cdot M_{\text{tot}} \cdot r_{\rm peri}}{V_{\infty}^2}}
\end{equation}
We set the semi-major axis of the binary, $a$, as the pericenter between the binary and the single in order to get stronger interactions that are more likely to result in resonant flybys, exchanges or mergers.

For each semi-major axis bin, we pick 100 impact parameters from a uniform distribution between 0 and b$_{\text{max}}^2$ and take the square root of these values. For each interaction we also assign three different seeds, this results in a total of 800 unique interactions and 2400 interactions in total. We create two similar data sets: one where we start with the star and a BH in the binary (test setup 1) and one with the two BHs in the binary (test setup 2). All initial parameters are the same for both sets and the only difference is whether the star is initially in the binary or as the single incoming object.

\begin{table}
    \caption{Initial properties for the test setup. For the semi-major axis we create 8 bins equally distributed in log between 5 R$_{\odot}$ and 1 AU. For each semi-major axis bin we calculate the maximum b$_{\text{max}}$ and pick 100 values from a uniform distribution between 0 and b$_{\text{max}}$.}
    \begin{tabular}{ll}
        \toprule
        Property & Value \\
        \midrule
        M$_{\star}$ & 1 M$_{\odot}$ \\
        M$_{BH}$ & 10 M$_{\odot}$   \\
        a & Min: 5 R$_{\odot}$, Max: 1 AU \\
        e & 0 \\
        b & Min: 0, Max: b$_{\text{max}}$ (see eq.\ref{eq:bMax}) \\
        V$_{\infty}$ & 5 km/s \\
        \bottomrule
    \end{tabular}
    \label{tab:testSetup_initalProperties}
\end{table}

\subsection{Results}
The outcomes from the test setup are shown in tables \ref{tab:manual_BHStar}, \ref{tab:manual_BHStar_mergers} and \ref{tab:manual_BHBH}. Table \ref{tab:manual_BHStar} and \ref{tab:manual_BHStar_mergers} shows the outcomes when the initial binary contains a BH and a star (test setup 1) while table \ref{tab:manual_BHBH} shows the outcomes when the initial binary contains two BHs (test setup 2). Some cells contains two rows, here the first number shows the fraction of interactions that ends with this outcome relative to the total number of interactions and the second number shows the total number of interactions with this outcome. The number inside the brackets indicate the number of unique interactions, i.e. identical interactions with the same outcome but with different seeds are only counted once. Looking at table \ref{tab:manual_BHStar} we can see that for all outcomes; flybys, exchanges and mergers, we see no significant differences between the no flags and PN terms run. However, when including tidal effects we see an increase in mergers and a drop in flybys and exchanges as a result of this. In table \ref{tab:manual_BHStar_mergers} we take a closer look at the mergers from table \ref{tab:manual_BHStar}. As previously mentioned the number of mergers increase when tides are involved. Since the cross-section for a merger with tides is higher than without, this is expected. Looking further into these mergers we can see that with no flags and only tidal effects we have no BH$_1$:BH$_2$ mergers, however, with PN terms and PN + tides we have one merger for each run. As expected it is the number of BH-star mergers that is increased by tides which increase by approximately 54 per cent when tidal effects are included. We can split these mergers into mergers between the initial single incoming BH (BH$_1$) and the initial binary BH (BH$_2$). The increase in the number of BH$_2$:S mergers ($\sim$75 per cent) is noticeably higher than for BH$_1$:S mergers ($\sim$35 per cent) when comparing the number of mergers in the runs without tides to the runs with tidal effects. A reason behind this might be that the high mass incoming BH strongly perturbs the orbit of the star to the point where it merges with the binary companion BH.

\begin{figure*}
    \centering    \subfigure[]{\includegraphics[width=0.49\textwidth]{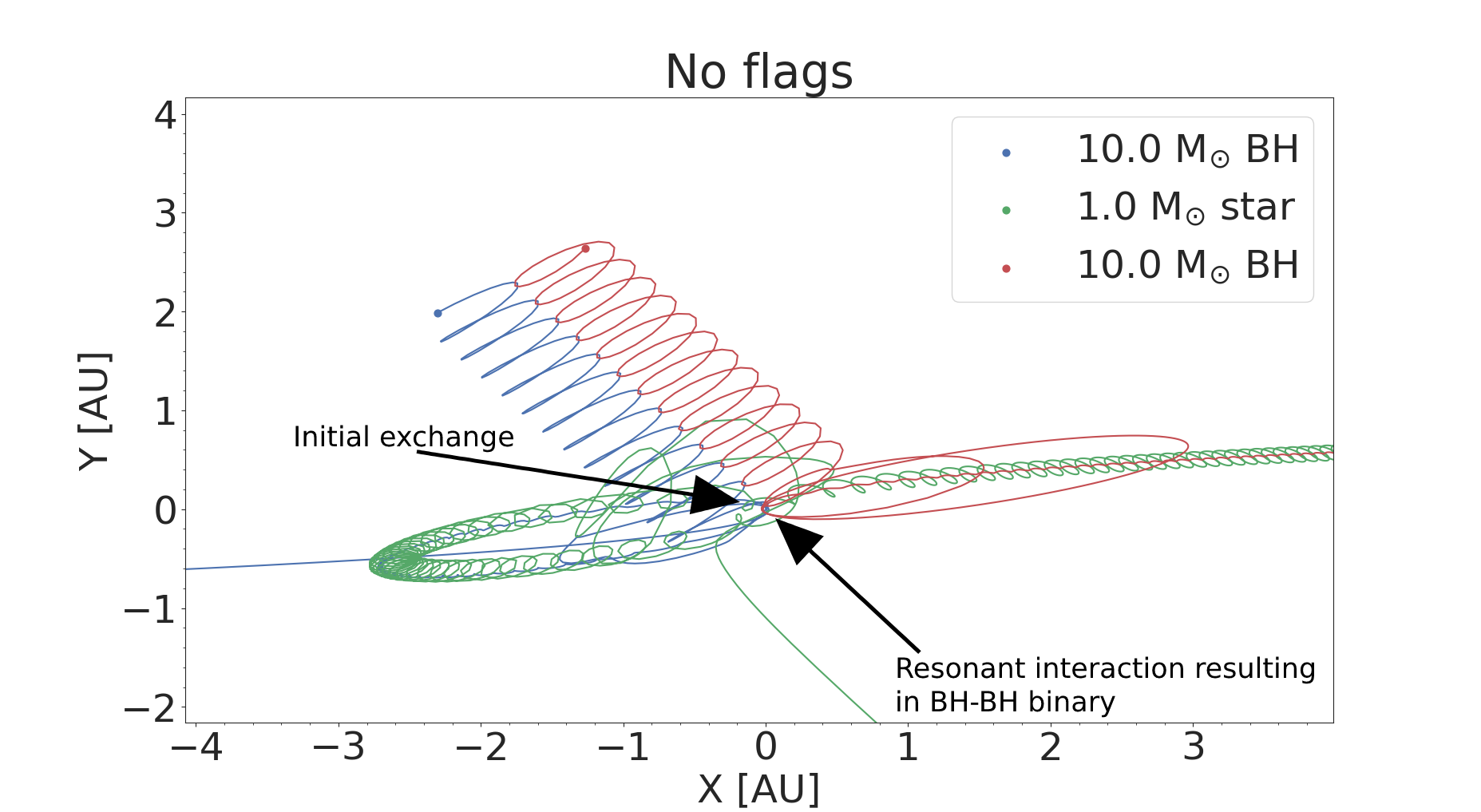}} 
    \subfigure[]{\includegraphics[width=0.49\textwidth]{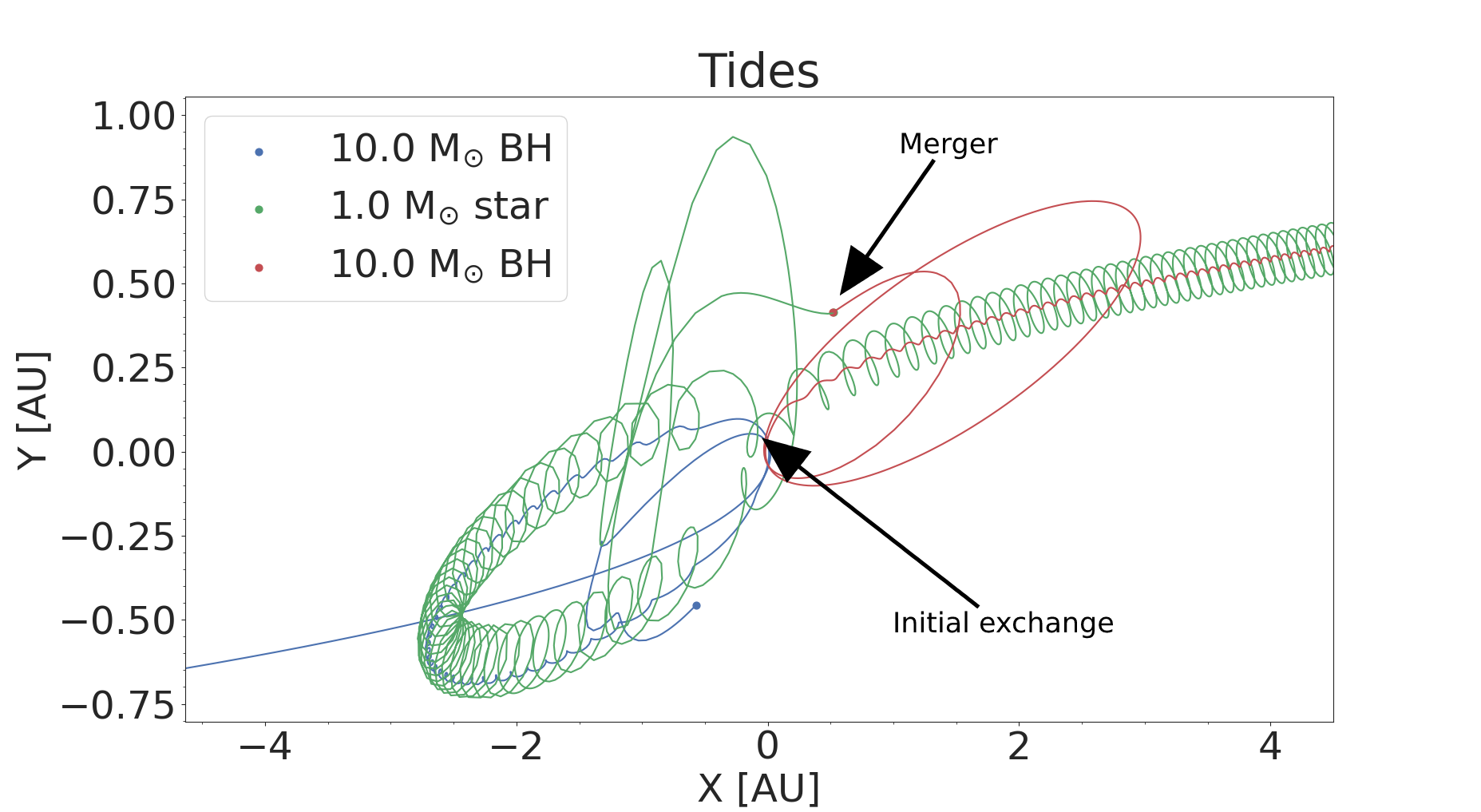}} 
    \caption{X-Y plane trajectories for the no flags (a) and tides (b) run of the example from the test setup. The initial setup is BH$_1$ [BH$_2$ S]. Both runs are resonant but the outcome is not the same. The no flags run ends with an exchange and a BH-BH binary while the tides run ends in a BH-Star merger between the two initial binary components. Note that the scales of the figures are different in order to show the important details of both interactions.}
    \label{fig:testSetup_example}
\end{figure*}

\begin{table}
    \caption{Outcomes for Test setup 1: The first column shows which flags are used, the second columns shows the total number of interactions in the run, the third column shows the number of flybys, the fourth column shows the number of exchanges and the fifth columns shows the number of mergers. The number in brackets represents the number of `unique' interactions with this outcome, i.e. when we only count each interaction once if different seeds give the same outcome. Each row for the different outcomes first shows the fraction of interactions with this outcome relative to the total number of interactions in the set and below it the total number of interactions with this outcome.}  
    \begin{tabular}{lcccccccr}
		\toprule
                 & Total & Flybys & Exchanges   & Mergers   \\
        		\midrule
	    \multirow{2}{*}{No flags} & \multirow{2}{*}{2398 (800)} & 0.138     & 0.728       & 0.133        \\
                                   &                              & 332 (284) & 1746 (777)  & 320  (243)   \\\midrule
        \multirow{2}{*}{PN}       & \multirow{2}{*}{2398 (800)} & 0.136     & 0.723       & 0.141        \\
                                   &                              & 327 (275) & 1733 (770)  & 338  (253)   \\\midrule
        \multirow{2}{*}{Tides}    & \multirow{2}{*}{2396 (800)} & 0.124     & 0.659       & 0.217        \\
                                   &                              & 297 (254) & 1579 (747)  & 520 (358)   \\\midrule
        \multirow{2}{*}{PN+Tides} & \multirow{2}{*}{2397 (800)} & 0.126     & 0.669       & 0.205        \\
                                   &                              & 303 (252) & 1603 (748)  & 491 (339)   \\
        \bottomrule
    \end{tabular}
    \label{tab:manual_BHStar}
\end{table}

\begin{table}
    \caption{Table gives details of the mergers seen in Test setup 1. The first column shows which flags are used, the second column shows the total number of mergers, the third column shows the number of BH-BH mergers, the fourth column shows the number of mergers between any of the BHs and the star. The fifth and sixth columns shows the number of mergers between the single BH and star and between the initial binary BH and the star respectively. The rows have the same description as provided in the caption of table \ref{tab:manual_BHStar}.}  
    \begin{tabular}{lcccccccr}
		\toprule
                 & Mergers   & BH$_1$:BH$_2$ & BH$_1$:S & BH$_2$:S \\
        		\midrule
        \multirow{2}{*}{No flags} &  \multirow{2}{*}{320  (243)}  & 0       & 0.547      & 0.453            \\
                                   &                                & 0       & 175 (150)  & 145 (126)        \\\midrule
                                   
        \multirow{2}{*}{PN}       & \multirow{2}{*}{338  (253)}  & 0.003    & 0.527      & 0.470            \\
                                   &                               & 1 (1)    & 178 (150)  & 159 (142)        \\ \midrule
                                   
        \multirow{2}{*}{Tides}    & \multirow{2}{*}{520  (358)}  & 0       & 0.463      & 0.537             \\
                                   &                               & 0       & 241 (203)  & 279 (222)         \\\midrule
                                   
        \multirow{2}{*}{PN+Tides} & \multirow{2}{*}{491  (339)}  & 0.002    & 0.466      & 0.532            \\
                                   &                               & 1 (1)    & 229 (192)  & 261 (210)        \\
        \bottomrule
    \end{tabular}
    \label{tab:manual_BHStar_mergers}
\end{table}

In table \ref{tab:manual_BHBH} we show the outcomes for interactions where the initial binary contains two BHs (test setup 2). In this set we see no exchanges since it is generally unfavourable to exchange in an object with 10 times lower mass than any of the binary members. A majority of the interactions in all runs results in a flyby, however, we have more mergers ($\sim$35 per cent increase) when including tides. We do not have any BH$_1$:BH$_2$ mergers so all of this increase is found in the BH:S mergers.

\begin{table}
    \caption{Outcomes for Test setup 2: The first columns shows which flags are used the second column shows the number of flybys and the third columns shows the number of mergers between any of the BHs and the star. No interactions in this set ends with an exchange or BH-BH mergers. The rows have the same description as provided in the caption of table \ref{tab:manual_BHStar}.}
    \begin{tabular}{lcccccr}
        \toprule
             & Total & Flybys & BH$_{1,2}$:S \\
             \midrule
        \multirow{2}{*}{No flags}  & \multirow{2}{*}{2400 (800)} & 0.813      & 0.188            \\
                                    &                              & 1950 (774) & 450 (313)        \\ \midrule
                                    
        \multirow{2}{*}{PN}        & \multirow{2}{*}{2398 (800)} & 0.821      & 0.179            \\
                                    &                              & 1968 (779) & 430 (308)        \\ \midrule
                                    
        \multirow{2}{*}{Tides}     & \multirow{2}{*}{2399 (800)} & 0.757      & 0.243            \\
                                    &                              & 1815 (755) & 584 (391)        \\ \midrule
                                    
        \multirow{2}{*}{PN+Tides}  & \multirow{2}{*}{2397 (800)} & 0.752      & 0.248            \\
                                    &                              & 1802 (752) & 595 (384)        \\
        \bottomrule
    \end{tabular}
    \label{tab:manual_BHBH}
\end{table}

\subsection{Example interaction}
\label{sec:testSet_exampleInt}
Here we present an example of a BH$_1$ [BH$_2$ S] interaction from the test setup where the inclusion of tides changes the outcome of the interaction. All initial parameters are constant in the test setup except for the semi-major axis and impact parameter. The parameters are shown in table \ref{tab:testSetup_initalProperties} with semi-major axis $a=0.117$ AU and impact parameter $b=8.6$ AU.

The trajectories in the X-Y plane for this interaction are shown in figure \ref{fig:testSetup_example}. Panel (a) shows the trajectory for an interaction with no additional processes included, for this run we initially have an exchange where we form a temporary triple with the inner binary consisting of the incoming BH and the star with the initial binary BH as the third outer object. The three objects quickly return to pericenter and we have a chaotic interaction which eventually ends with the formation of a BH-BH binary and the star as a single object. In panel (b) we show the trajectory with tides included, in this case we have the same initial exchange and form a temporary bound triple with the incoming BH and the star as the inner binary. When the three objects return to pericenter we have a chaotic encounter which ends with a merger between the two initial binary components.

The resulting binary in the no flags run have properties $a=0.6552$ AU, $e=0.905$ and M$_{1}=$ M$_{2}=10$ M$_{\odot}$. Using equation \ref{eq:GWMergers} we can get an estimate for the merger time of this binary due to gravitational wave radiation: $t_{GW} \approx 7.08 \cdot 10^{9}$ yrs. For the tides run we use the sticky-star approximation to get the state vector of the centre of mass at the point of the merger. We assume no mass loss and conserve linear momentum during the merger, and use the centre of mass as the remaining object. Calculating the orbital parameters for the orbit between the merger product and the third object we get  $a=0.953$ AU and  $e = 0.9662$, with the masses M$_1 = 10$ M$_{\odot}$ and M$_{2} = 11$ M$_{\odot}$. This gives a merger time of $t_{GW} \approx 1.10 \cdot 10^{9}$ yrs. From this we can see that, for this example, including tides not only leads to a BH-Star merger but also the final binary will have a lower merger time compared to the binary that's created through the exchange in the run with no additional processes.

\begin{figure*}
    \centering
    \includegraphics[width=\textwidth]{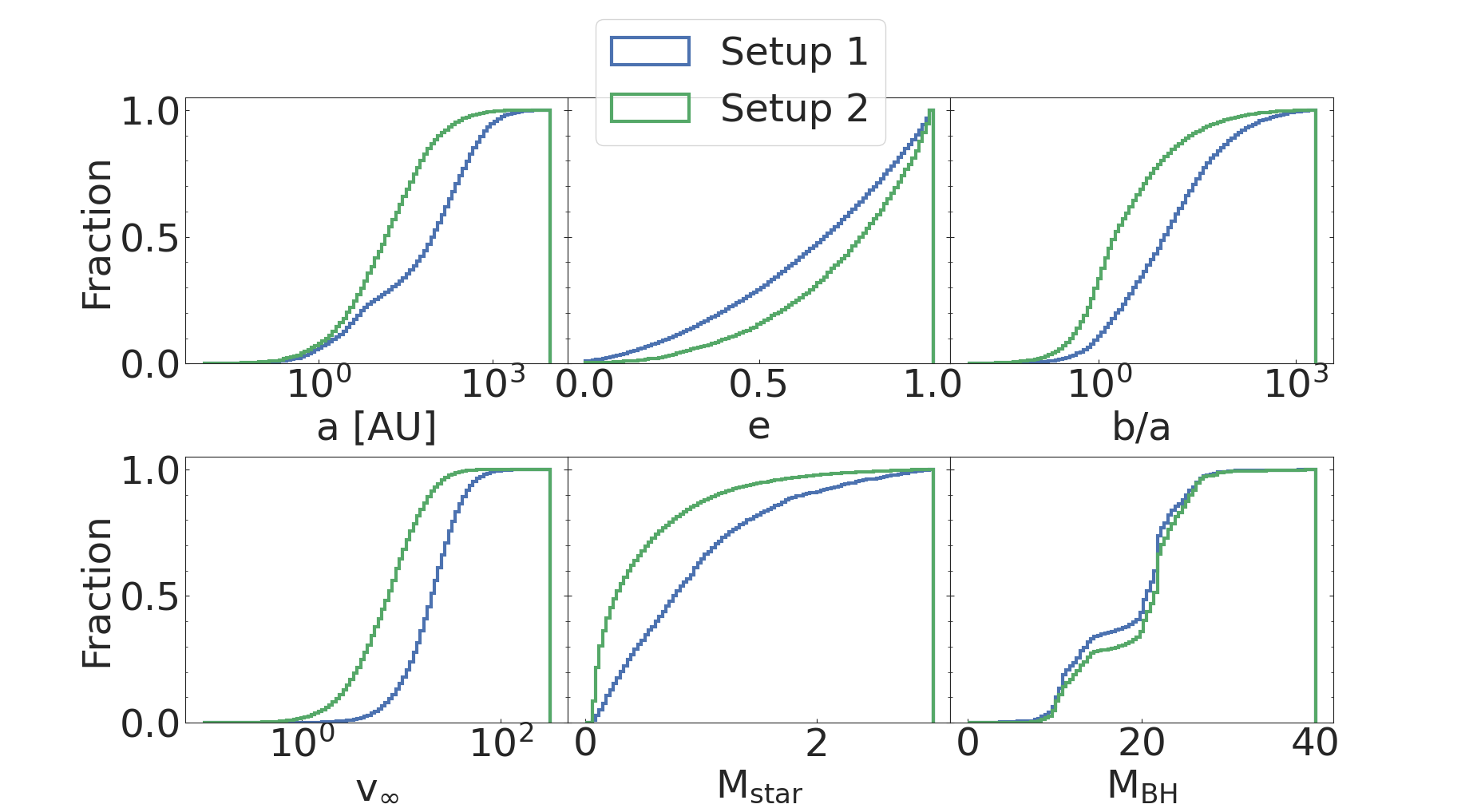}
    \caption{Cumulative distribution plots for the initial properties in the M1 set (from left to right); semi-major axis, eccentricity, impact parameter, velocity at infinity, star mass and BH mass. Setup 1 is the case where the star is initially in the binary and the incoming object is a BH. Setup 2 is the case where the initial binary contains two BHs and the incoming object is a star.}
    \label{fig:M1_initialProp}
\end{figure*}

\section{Data from MOCCA}
\label{sec:moccaSetup}
\subsection{MOCCA set 1 (M1)}
\label{sec:M1}
We extract approximately 82 000 binary-single interactions with two BHs and a star from several different MOCCA simulations of globular cluster evolution. These interactions are simulated using \textsc{tsunami}  with a seed picked from a uniform random distribution which results in different initial orientations for all interactions. All of these interactions are simulated using one seed for each interaction.

\subsubsection{Initial properties}
\label{sec:M1_initParams}
The initial parameters for set M1 are presented in figure \ref{fig:M1_initialProp}. Each panel is split into two cumulative distributions, one for interactions with an initial BH-BH binary (M1 setup 1) and one for interactions with an initial BH-Star binary (M1 setup 2). From left to right, the first panel shows the semi-major axis distribution which ranges from 10$^{-1}$ AU to 10$^{4}$ AU. Setup 1 have, on average, higher semi-major axis than the setup 2. This might be connected to the way binaries are formed in three-body interactions in MOCCA. The semi-major axis is proportional to three times the average kinetic energy and thus if the binary components are very massive the semi-major axis can be very large. We can also see that in setup 1 there is a small bump at around 8 AU. This bump is related to primordial BH-BH binaries while most of the rest of the curve is dominated by dynamically formed BH binaries.

The eccentricity is shown in the next panel where setup 1 have, on average, lower eccentricity than setup 2. A reason for this might be that many of the BH-Star binaries form dynamically and are frequently perturbed by surrounding single BHs and flyby interactions which can increase their eccentricity. We also have a few binaries with 0 eccentricity. The BH-Star binaries with 0 eccentricity are most likely circularised by tidal forces \citep{hurley2002} while for the BH-BH binaries, gravitational wave radiation can circularise the system \citep{gravRadMergerTime}. The distribution of eccentricities in a relaxed system is expected to be thermal \citep{jeansThermalEcc, HeggieBinaryEvolution,leigh2016}. However, we can see that our distribution deviates from a thermal distribution and we have many more higher eccentricity binaries. This is probably because such binaries are subject to frequent flyby interactions that can pump up their eccentricity. Additionally, the interactions that we extract originate from multiple clusters at different times during their evolution.

In the next panel, the impact parameter is shown, we can see that setup 2 have, on average, lower impact parameter in relation to the semi-major axis compared to setup 1. This is caused by the differences in semi-major axis. The first panel on the second row shows the initial velocity at infinity where we can see that setup 1 have, on average, higher v$_{\infty}$ than setup 2. In a relaxed system such as a globular cluster, we evolve towards energy equipartition. By assuming that the stars have masses proportional to the average mass in the cluster we can say that the average velocity for a star is close to the average velocity in the cluster. Since the average mass of a BH is larger than the average mass of regular, the average velocity of BHs is also lower. Thus, the average relative velocity between a star and a BH binary is larger than the one between a BH and a BH-star binary.

The middle panel on the second row shows the masses of stars for the interactions with both initial configurations; setup 1 and 2. We can see that single stars have on average lower mass than the stars found in BH-Star binaries in this set. A reason for this might be that it is difficult to have primordial binaries with very massive and low mass components. A binary between a massive BH and a star is also likely to be broken up by dynamical interactions. We can also assume that when a BH binary interacts with a star, the mass of the star follows the distribution of single stars in the system. However, stars in binaries has to be more massive than single stars since most BH-star binaries are formed in exchanges where usually the more massive star is bound to the BH. The final panel shows the BH masses and here the BH masses are very similar and the differences between the incoming BHs and the binary component BHs are very small.

The masses reflect the BH mass distribution connected with the assumed prescriptions for the evolution of massive stars in the \textsc{MOCCA}-Survey Database I \citep{askar2017} simulations. For binary/stellar evolution, these runs used prescriptions from \citet{hurley2002}. In some of the runs, the BH mass is modified according to the mass fallback prescriptions provided by \citet{belczynski2002}. The maximum BH mass produced in through the evolution of a single star was about $\rm 30 M_{\odot}$. We can see from both distributions that there seems to be two groups of BHs, one group where the mass is between 10 and 15 M$_{\odot}$, which most likely corresponds to BHs without mass fallback and one group with mass > 20 M$_{\odot}$, which most likely corresponds to BHs with mass fallback \citep{massFallbackZhang}. For the evolution of BH progenitors, these simulations did not consider metalliticy dependent winds, pair/pulsational pair instability supernovae and the supernova explosion model \citep{fryer2012,woosley2017,spera2017,vink2021}.

\begin{table}
    \caption{M1 setup 1 outcomes: The first column shows which flags were used, the second column shows the total number of completed interactions. The remaining columns shows the number of flybys, exchanges, mergers and Ionisations respectively. The rows have the same description as provided in the caption of table \ref{tab:manual_BHStar}.}
    \begin{tabular}{lcccccccr}
    \toprule
         & Total & Flybys & Exchanges   & Mergers   & Ionisations \\
         \midrule
    \multirow{2}{*}{No flags} & \multirow{2}{*}{5148} & 0.645 & 0.269 & 0.018 & 0.068   \\
                              &                        & 3321  & 1382  & 94    & 351     \\ \midrule
                               
    \multirow{2}{*}{PN}       & \multirow{2}{*}{5143} & 0.643 & 0.269 & 0.020 & 0.068   \\
                              &                        & 3309  & 1381  & 102   & 351     \\ \midrule
                               
    \multirow{2}{*}{Tides}    & \multirow{2}{*}{5057} & 0.647 & 0.257 & 0.028 & 0.068   \\
                              &                       & 3270  & 1302  & 140   & 345     \\ \midrule
                               
    \multirow{2}{*}{PN+Tides} & \multirow{2}{*}{5051} & 0.645 & 0.255   & 0.031 & 0.069   \\
                              &                       & 3260  & 1289  & 156   & 346     \\
    \bottomrule
    \end{tabular}
    \label{tab:M1_BHStar}
\end{table}

\begin{table}
    \caption{M1 setup 1 mergers: The first column shows which flags were used, the second column shows the total number of mergers and the rest of the columns shows the number of BH-BH mergers, mergers between the incoming BH and the star and the last column shows the number of mergers between the initial binary components. The rows have the same description as provided in the caption of table \ref{tab:manual_BHStar}.}
    \begin{tabular}{lcccccccr}
    \toprule
         & Mergers   & BH$_1$:BH$_2$  & BH$_1$:S & BH$_2$:S \\
         \midrule
    \multirow{2}{*}{No flags} & \multirow{2}{*}{94}    & 0       & 0.351  & 0.649     \\
                              &                        & 0       & 33     & 61        \\ \midrule
                               
    \multirow{2}{*}{PN}       & \multirow{2}{*}{102}   & 0.029   & 0.314  & 0.657     \\
                              &                        & 3       & 32     & 67        \\ \midrule
                               
    \multirow{2}{*}{Tides}    & \multirow{2}{*}{140}   & 0       & 0.450  & 0.55      \\
                              &                        & 0       & 63     & 77        \\ \midrule
                               
    \multirow{2}{*}{PN+Tides} & \multirow{2}{*}{156}   & 0.013   & 0.487  & 0.50      \\
                              &                        & 2       & 76     & 78        \\ 
    \bottomrule
    \end{tabular}
    \label{tab:M1_BHStar_mergers}
\end{table}

\begin{table*}
    \caption{M1 setup 2 outcomes: The first column shows which flags were used, the second column shows the total number of complete interactions and the third column shows the number of flybys. The fourth column shows the number of exchanges, the fifth column shows the total number of mergers. The sixth and seventh column shows the number of BH-BH mergers and BH-Star mergers respectively. The eighth column shows the number of Ionisations and the ninth and final column shows the number of stable triples. The rows have the same description as provided in the caption of table \ref{tab:manual_BHStar}.}
    \begin{tabular}{lcccccccc}
        \toprule
                 & Total & Flybys & Exchanges & Mergers    & BH$_1$:BH$_2$ & BH$_{1,2}$:S & Ionisations & Stable triples \\
                 \midrule
        \multirow{2}{*}{No flags} & \multirow{2}{*}{76842} & 0.9897 & 3e-4   & 0.01  & 0      & 0.01   & 9e-5 & 0       \\
                                   &                       & 76048  & 23     & 758   & 0      & 758    & 7    & 0       \\ \midrule
                                   
        \multirow{2}{*}{PN}       & \multirow{2}{*}{76838} & 0.9898 & 4.1e-4 & 0.01  & 2.6e-5 & 0.01   & 9e-5 & 1.3e-5  \\
                                   &                       & 76055  & 32     & 744   & 2      & 742    & 7    & 1       \\ \midrule
                                   
        \multirow{2}{*}{Tides}    & \multirow{2}{*}{76833} & 0.976  & 3e-4   & 0.023 & 0      & 0.023  & 9e-5 & 0       \\
                                   &                       & 74994  & 23     & 1803  & 0      & 1803   & 7    & 0       \\ \midrule
                                   
        \multirow{2}{*}{PN+Tides} & \multirow{2}{*}{76818} & 0.976  & 3.9e-4 & 0.023 & 2.6e-5 & 0.023  & 9e-5 & 0       \\
                                   &                       & 74984  & 30     & 1797  & 2      & 1795   & 7    & 0       \\
        \bottomrule
    \end{tabular}
    \label{tab:M1_BHBH}
\end{table*}

\subsubsection{Results}

Tables \ref{tab:M1_BHStar}, \ref{tab:M1_BHStar_mergers} and \ref{tab:M1_BHBH} show the outcomes for set M1. Table \ref{tab:M1_BHStar} shows all outcomes for interactions from M1 setup 1, we can see that a majority of interactions result in a flyby followed by exchanges. The number of mergers is small but we can see that the inclusion of tides increase the merger rate by approximately 40 per cent. We also have a few ionisations where the incoming single causes the binary to ionise and the end result is three single objects. In table \ref{tab:M1_BHStar_mergers} we take a closer look at the number of mergers. Without PN terms we do not see any BH:BH mergers, however, with PN terms included we see 3 for the run with only PN terms and 2 for the run with PN terms and tides. The inclusion of tides leads to an increase of approximately 40 per cent for the BH:S mergers. The merger rates between BH$_1$ and the star increases by approximately 100 per cent while the merger rates between the two initial binary components increase by approximately 26 per cent. Compared to the Test setup (see sec \ref{sec:testSetup}), the tidal effects increase the number of BH$_1$:star mergers to a much larger extent in M1 while for the Test setup, the number of BH$_2$:star mergers see the largest increase. The reason for this might be that the binaries found in MOCCA are more stable and require stronger perturbations for the two binary components to merge. The binaries from MOCCA have also, on average, higher semi-major axis than the binaries we setup in the test setup. The masses are also very different; in the test setup the BH masses are always 10 times higher than the star mass. In the MOCCA set this varies from interaction to interaction.

\begin{figure*}
    \centering
    \includegraphics[width=\textwidth]{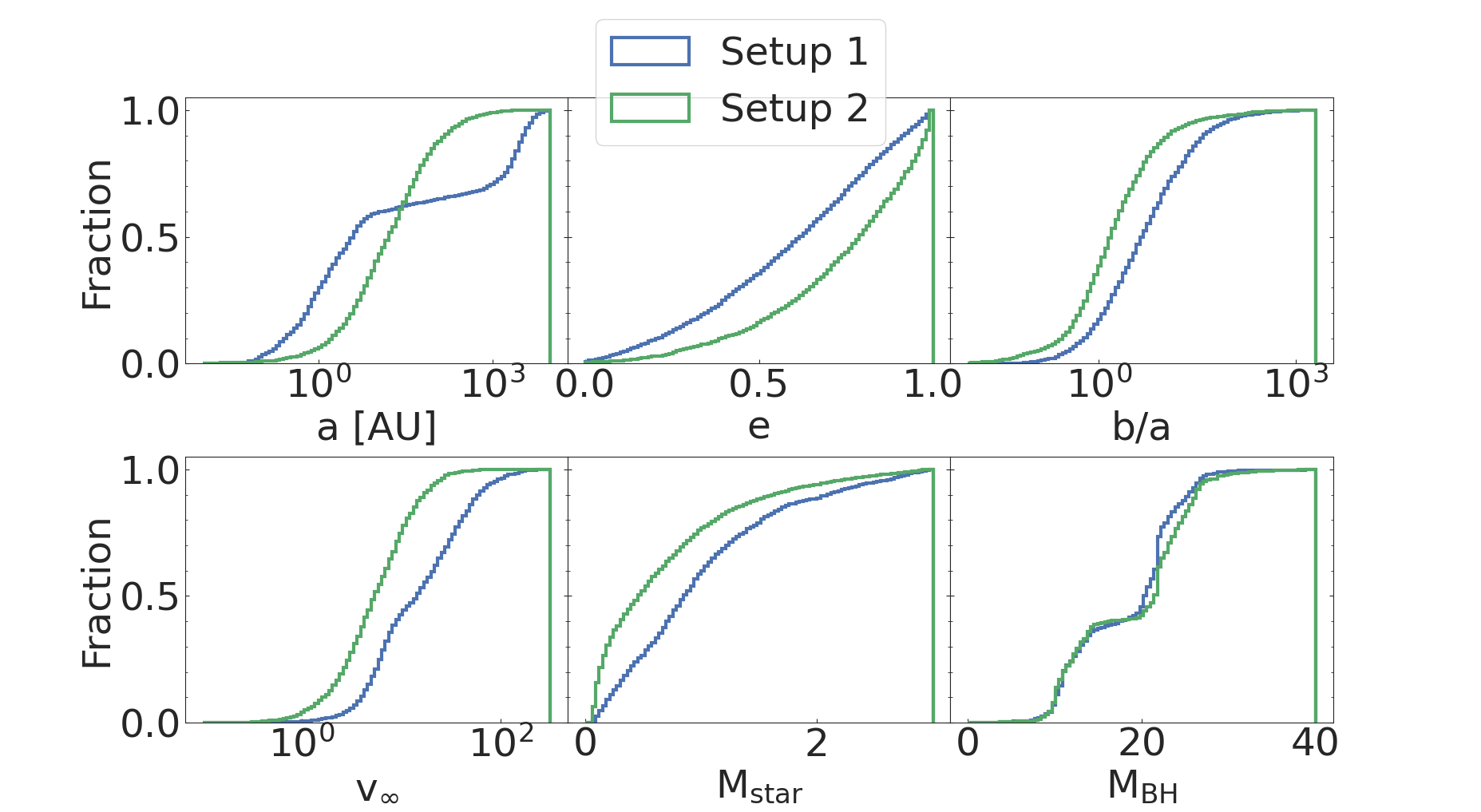}
    \caption{Cumulative distribution plots for initial properties distributions in the M2 set (from left to right); semi-major axis, eccentricity, impact parameter, velocity at infinity, star mass and BH mass.}
    \label{fig:M2_initialProp}
\end{figure*}

In table \ref{tab:M1_BHBH} we show the number of outcomes for the interactions from M1 setup 2. In this data set, a large majority of the outcomes results in flybys and a small fraction as exchanges. The number of mergers is low compared to the total number of interactions, however, if we compare the runs with and without tides we can see that the inclusion of tides increase the number of BH:S mergers by approximately 140 per cent. BH:BH mergers are only found in the two runs with PN terms and the large increase in mergers are due to an increase in BH:S mergers. We also have a very small amount of ionisations, however, this number is consistent for all four runs. With PN terms we see one stable hierarchical triple where the BH binary captures the star in a wide orbit around the binary. \textsc{tsunami}  uses the triple stable criterion described in section \ref{sec:tripleStability} in order to classify hierarchical systems as stable. The fact that we only see one of this outcome means that, in our data set and with this classification criteria, it is very unlikely to form stable hierarchical triples.

\subsection{MOCCA set 2 (M2)}
\label{sec:M2}
\subsubsection{Initial parameters}
A large majority of the interactions in MOCCA set 1 ends as a flyby and the interactions are quite distant, weak and non-resonant. In order to get a better understanding of how the additional processes affects stronger interactions we create a filtered MOCCA set called MOCCA set 2 (M2). This set is created by integrating all interactions in M1 without any additional processes and a selected seed such that the interaction is co-planar. From these runs, all non-flybys are extracted and assigned 5 seeds for each interaction which results in a set of 5.6k unique interactions and 28k total interactions. We split the interactions into two groups; interactions where the star is initially in the binary and the incoming object is a BH (BH [BH S], setup 1) and interactions where the star is initially the incoming object and the binary is a BH-BH binary (S [BH BH], setup 2).

The initial parameters are shown in figure \ref{fig:M2_initialProp}. The initial semi-major axis distribution is found in the first panel on the upper row. We can see that on average, setup 2 have a lower semi-major axis than setup 1. We have two groups in setup 2, as discussed in M1, this is related to primordial and dynamically formed binaries. Primordial binaries have, on average, lower semi-major axis and correspond to the left-hand side of the curve while dynamically formed binaries correspond to the right-hand side. The second Panel shows the eccentricity distribution. The impact parameter distribution is shown in last panel of the upper row . The first panel on the lower row shows the velocity at infinity distribution. The second panel on the lower row shows the star mass and the final panel on the lower row shows the BH mass. Since this is a subset of M1, the initial parameter distributions are very similar, for explanations and discussions related to the initial parameters see section \ref{sec:M1_initParams}.

\begin{table*}
    \caption{M2 setup 1 outcomes: For a description of the columns, see table \ref{tab:M1_BHStar}. The rows have the same description as provided in the caption of table \ref{tab:manual_BHStar}.}
    \begin{tabular}{lccccccccr}
		\toprule
             & Total & Flybys & Exchanges   & Mergers   & Ionisations\\
    		\midrule
        \multirow{2}{*}{No flags} & \multirow{2}{*}{9703 (1945)} & 0.280       & 0.605       & 0.0106   & 0.105      \\
                                   &                               & 2713 (1172) & 5866 (1645) & 103 (79) & 1021 (474)           \\ \midrule
         
        \multirow{2}{*}{PN}       & \multirow{2}{*}{9693 (1945)} & 0.280       & 0.604       & 0.0103   & 0.105      \\
                                   &                               & 2718 (1166) & 5854 (1646) & 100 (78) & 1021 (474)           \\ \midrule
        
        \multirow{2}{*}{Tides}    & \multirow{2}{*}{9286 (1946)} & 0.263       & 0.610       & 0.0188   & 0.108    \\
                                   &                               &  2441 (1100) & 5668 (1596) & 175 (102) & 1002  (463)          \\ \midrule
        
        \multirow{2}{*}{PN+Tides} & \multirow{2}{*}{9287 (1946)} & 0.263       & 0.610       & 0.0188   & 0.108     \\ 
                                   &                               & 2447 (1093) & 5663 (1594) & 175 (99) & 1002  (463)           \\ 
		\bottomrule
    \end{tabular}
    \label{tab:M2_BHStar}
\end{table*}

\subsubsection{Results}

\begin{table}
    \caption{M2 setup 1 mergers: For a description of the columns, see table \ref{tab:M1_BHStar_mergers}. The rows have the same description as provided in the caption of table \ref{tab:manual_BHStar}.}
    \begin{tabular}{lccccccccr}
		\toprule
             & Mergers   & BH$_1$:BH$_2$  & BH$_1$:S & BH$_2$:S \\
    		\midrule
        \multirow{2}{*}{No flags} & \multirow{2}{*}{103 (79)}  & 0.049  & 0.107   & 0.845  \\
                                  &                            & 5 (1)  & 11 (10) & 87 (68)    \\ \midrule
         
        \multirow{2}{*}{PN}       & \multirow{2}{*}{100 (78)}  & 0.06   & 0.11    & 0.83 \\
                                  &                            & 6 (2)  & 11 (11) & 83 (66)    \\ \midrule
        
        \multirow{2}{*}{Tides}    & \multirow{2}{*}{175 (102)} & 0.029 & 0.12    & 0.85 \\
                                  &                            & 5 (1) & 21 (21) & 149 (91)     \\ \midrule
        
        \multirow{2}{*}{PN+Tides} & \multirow{2}{*}{175 (99)}  & 0.034 & 0.12    & 0.85  \\  
                                  &                            & 6 (2) & 21 (20) & 149 (88)     \\ 
		\bottomrule
    \end{tabular}
    \label{tab:M2_BHStarMergers}
\end{table}

This section will present the results from set M2 where we split up the interactions initially containing a BH-Star binary interacting with a single BH (M2 setup 1), shown in tables \ref{tab:M2_BHStar} and \ref{tab:M2_BHStarMergers}, and interactions initially containing a BH-BH binary interacting with a single star (M2 setup 2), shown in table \ref{tab:M2_BHBH}). 

We start by looking at the outcomes for interactions with a BH-Star binary (M2 setup 1) in table \ref{tab:M2_BHStar}. A large majority of these interactions results in either a flyby, an exchange or an ionisation. The number of flybys and exchanges decreases when tides are included but the number of ionisations are consistent between all 4 runs. The decrease in flybys and exchanges with tides included are due to the increase in mergers. Without tides we have 103 mergers in the no flags run and 100 mergers for the PN terms run while we have 175 mergers for both the run with tides and the run with PN terms and tides. This increase is significant, however, we have to consider that each interaction is run 5 time with different seeds. Thus we can look at the numbers in bracket and see that the increase in mergers is approximately 29 per cent which is more in line with the other two sets. By looking at these mergers in more depth in table \ref{tab:M2_BHStarMergers} we can see that the number of BH:BH mergers increase very slightly in the two runs where we use PN terms compared to the two runs where we do not. For the runs without PN terms we have 5 BH:BH mergers, but looking at the number of unique mergers we only have 1. The binary is very tight at $a=2.7 \cdot 10^{-2}$ AU and the impact parameter is very low at $b = 8.59 \cdot 10^{-3}$ AU. The single BH is travelling almost head-on and colliding very quickly with the binary component BH. Since the binary component have much higher mass than its binary companion star, the BH is always very close to the centre of mass and for all seeds that we have investigated this interaction leads to a merger. The two runs with PN terms have the same 5 BH:BH mergers as previously discussed but there is also one additional interaction where the two BHs merge for one seed only. 

The big increase with tides is, as expected, found in BH:S mergers where the increase in BH$_1$:S mergers is approximately 90 per cent both when considering the total and unique number of mergers. The increase in BH$_2$:S mergers is a bit less at approximately 71 per cent for the number of total mergers and 33 per cent for the number of unique mergers. We find that a few of these BH:S mergers happen shortly after the first close encounter with the single, this is most likely due to energy being transferred from the binary to the single which shrinks the binary to the point where the tidal effects cause an inspiral. These binaries that inspiral shortly after the interaction may explain the rest of the increase in number of mergers when tides are included.
\begin{table*}
    \caption{M2 setup 2 outcomes: For a description about the columns, see table \ref{tab:M1_BHBH}, with the exception that we do not find any stable triples in this setup and therefore do not include that column here. The rows have the same description as provided in the caption of table \ref{tab:manual_BHStar}.}
    \begin{tabular}{lccccccr}
        \toprule
             & Total & Flybys & Exchanges & Mergers & BH$_1$:BH$_1$ & BH$_{1,2}$:S & Ionisations \\
             \midrule
        \multirow{2}{*}{No flags} & \multirow{2}{*}{18176 (3638)}  & 0.98         & 6.6e-4   & 0.188       & 1.65e-4       & 0.0187     & 5.5e-5    \\
                                  &                                & 17821 (3638) & 12 (12)  & 342  (264)  & 3 (3)         & 339 (262)  & 1 (1)     \\ \midrule
                                  
        \multirow{2}{*}{PN}       & \multirow{2}{*}{18164 (3638)}  & 0.98         & 9.36e-4  & 0.188       & 3.30e-4       & 0.0187     & 5.5e-5    \\
                                  &                                & 17801 (3637) & 17 (15)  & 342  (278)  & 6 (2)         & 339 (276)  & 1 (1)     \\ \midrule
                                  
        \multirow{2}{*}{Tides}    & \multirow{2}{*}{18152 (3638)}  & 0.957        & 7.17e-4  & 0.042       & 1.65e-4       & 0.042      & 5.5e-5     \\
                                  &                                & 17371 (3636) & 13 (12)  & 767 (553)   & 3 (2)         & 764 (551)  & 1 (1)     \\ \midrule
                                  
        \multirow{2}{*}{PN+Tides} & \multirow{2}{*}{18158 (3638)}  & 0.956        & 7.16e-4  & 0.0436      & 3.86e-4       & 0.043      & 5.5e-5     \\
                                  &                                & 17353 (3635) & 13 (12)  & 791 (568)   & 7 (3)         & 784 (565)  & 1 (1)     \\
        \bottomrule
    \end{tabular}
    \label{tab:M2_BHBH}
\end{table*}

\begin{figure*}
    \centering
    \subfigure[]{\includegraphics[width=0.49\textwidth]{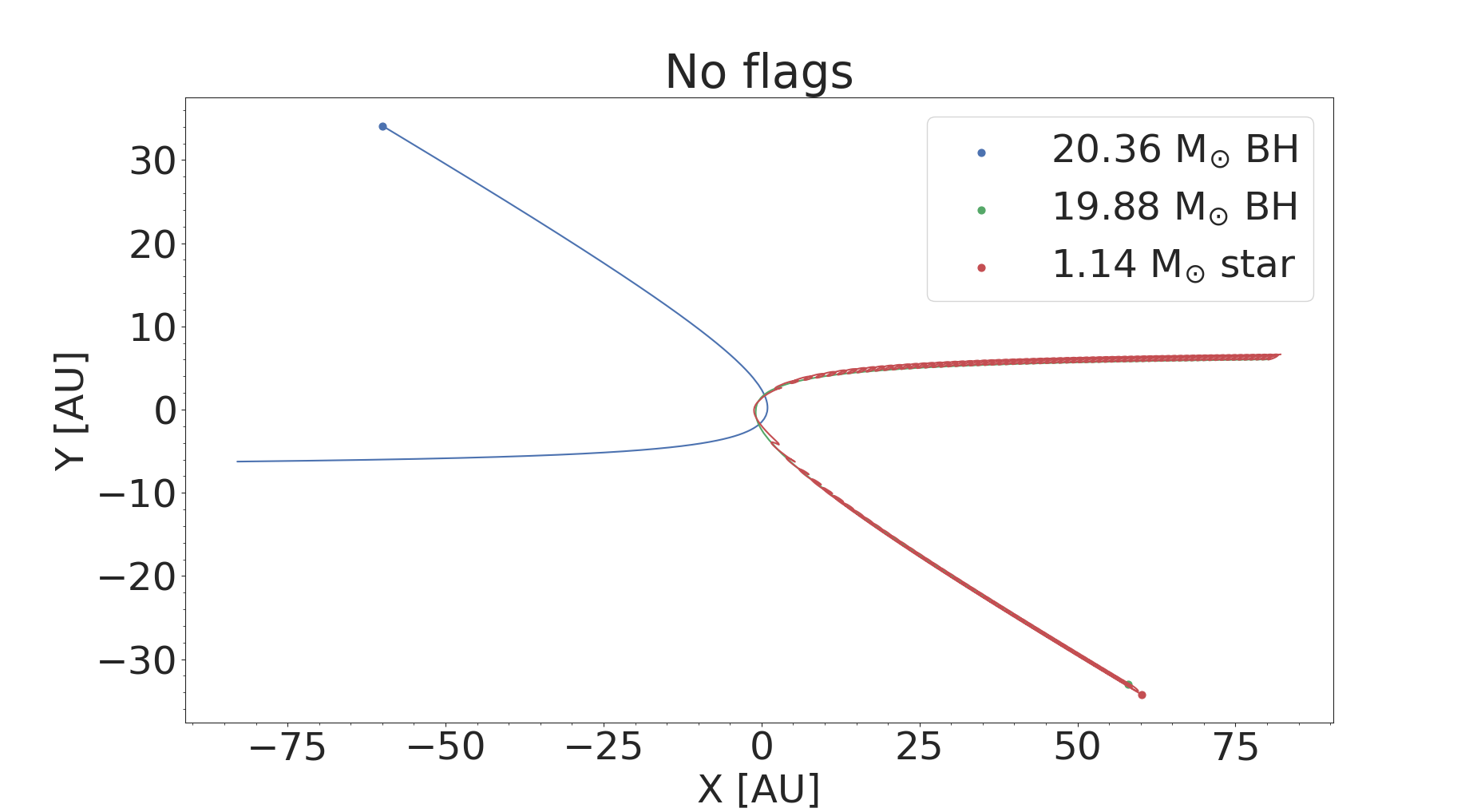}} 
    \subfigure[]{\includegraphics[width=0.49\textwidth]{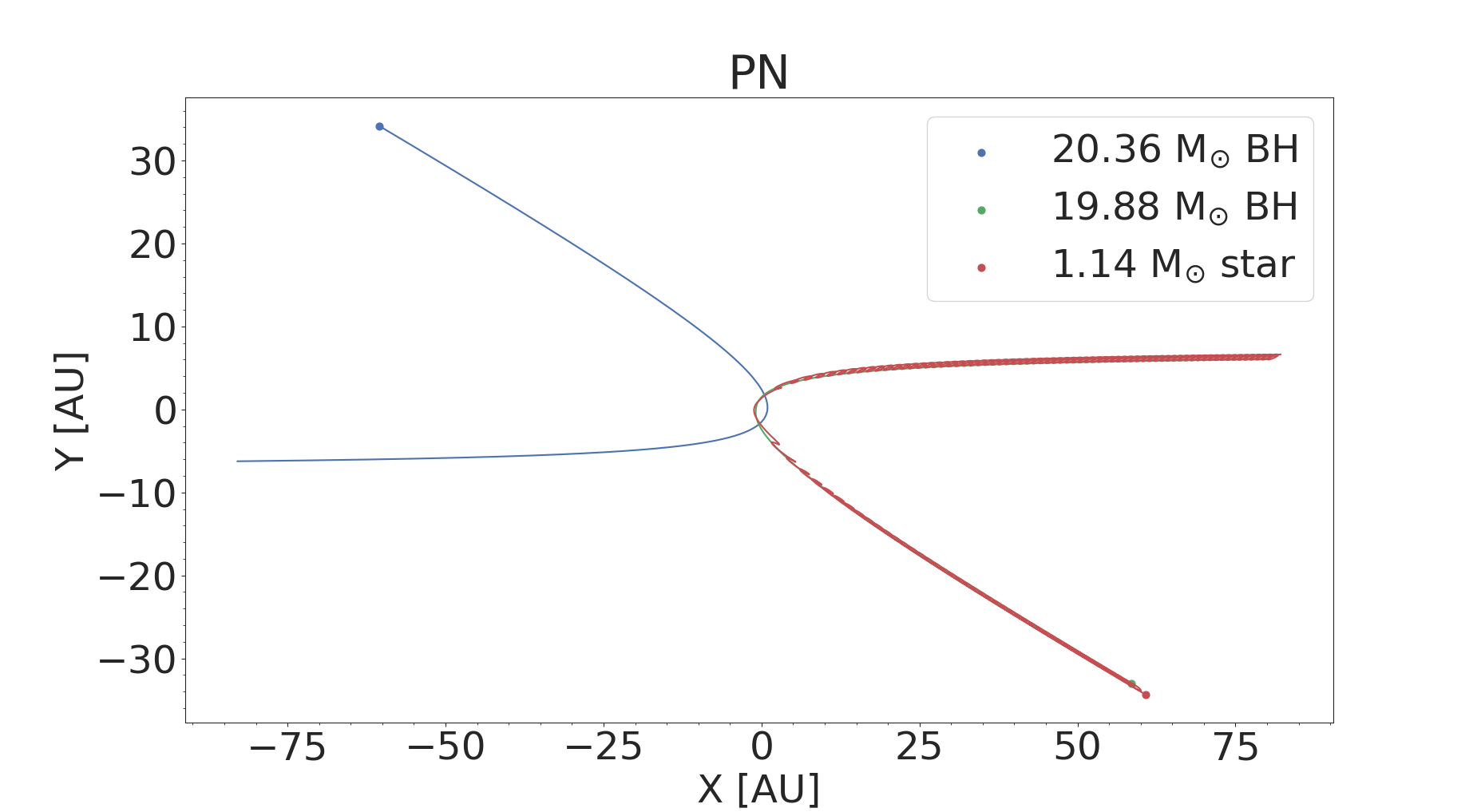}} 
    \subfigure[]{\includegraphics[width=0.49\textwidth]{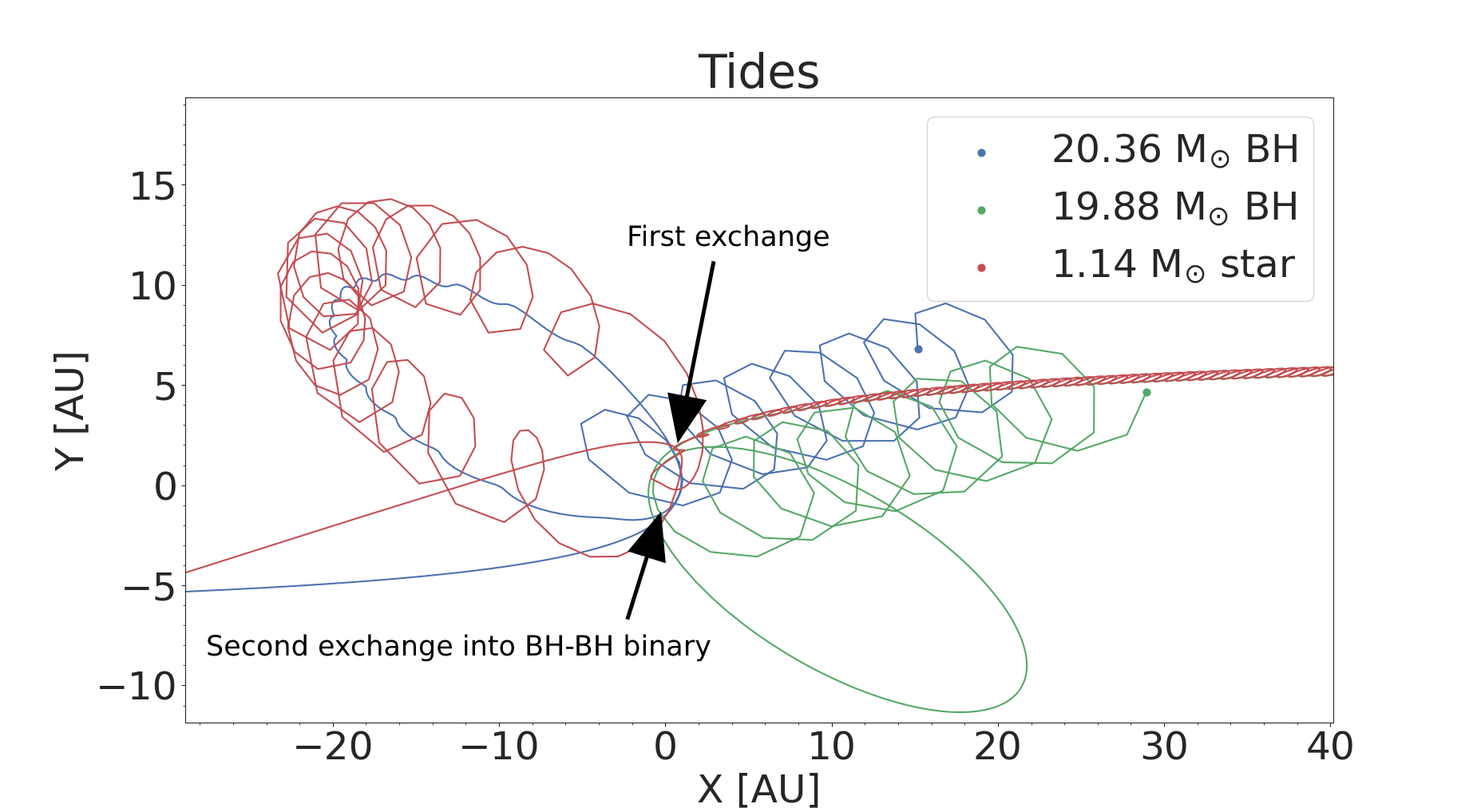}}
    \subfigure[]{\includegraphics[width=0.49\textwidth]{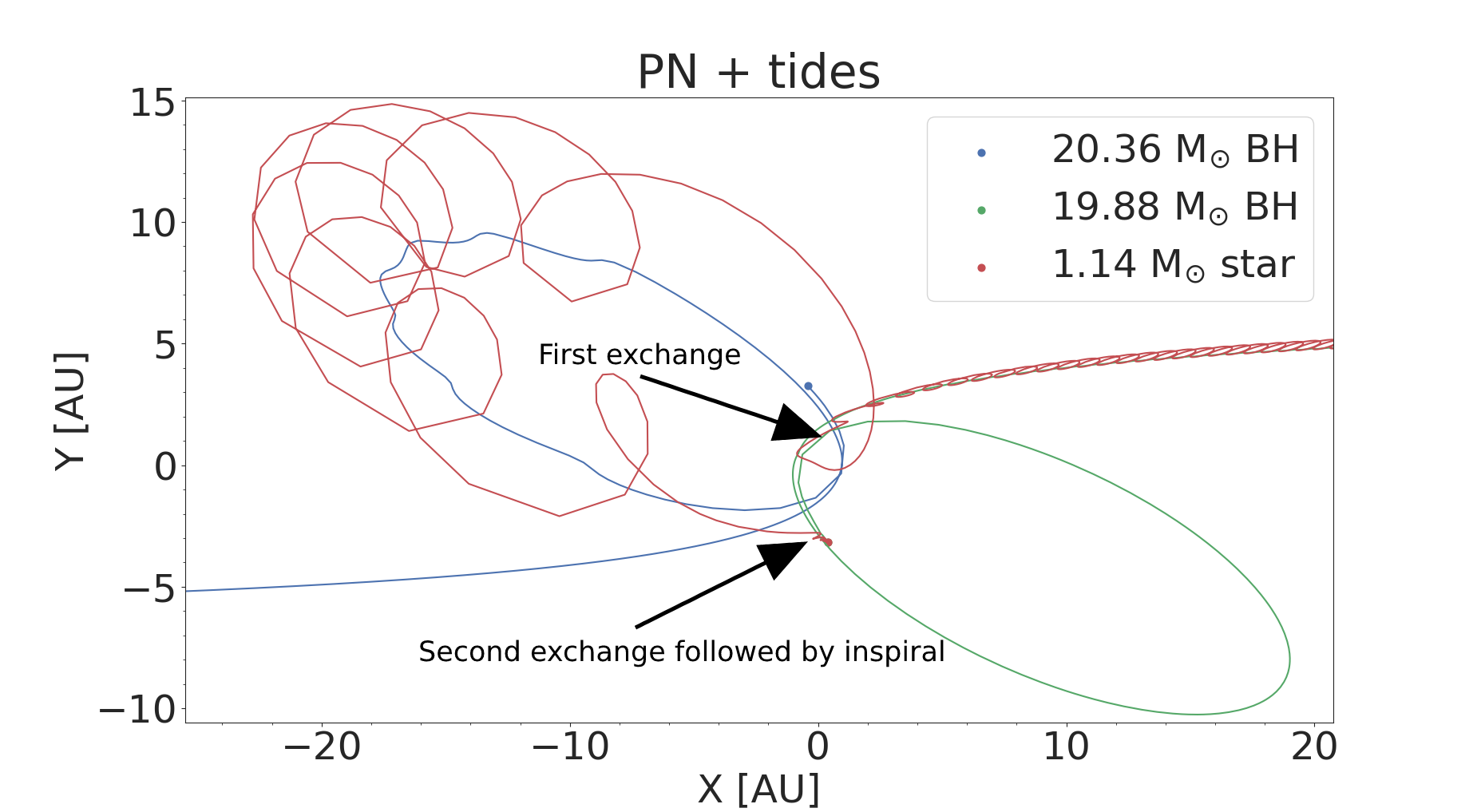}}
    \caption{X-Y plane trajectory for the interaction with no flags (a) which results in a flyby, PN (b) which also results in a flyby, tides (c) which results in an exchange and PN + tides (d) which results in a merger between the two initial binary components.}
    \label{fig:example_M2_traj}
\end{figure*}

The outcomes for M2 setup 2 are shown in table \ref{tab:M2_BHBH}. The majority of the interactions end as flybys, we have very few exchanges and the number of ionisations is constant for all four runs. The inclusion of tides increase the total number of mergers by approximately 124 per cent and the number of unique mergers by 109 per cent. The total number of BH:BH mergers increase by approximately 100 per cent when PN terms are included, however, the number of mergers are low and thus this may be due to statistical fluctuations. The change in the number of unique mergers is more difficult to quantify since the numbers are very low (two or three), however, from the results we have, we get one less unique merger with PN terms compared to no flags but one more merger with PN + tides compared to only tides. Thus, due to the extremely small sample size, it is not possible to draw any conclusions for this and the differences we see may be due to statistical fluctuations. 

The total number of BH:S mergers increase by approximately 125 per cent when including tides and the unique number of mergers increase by approximately 110 per cent. Compared to the other sets, this is the largest increase in the number of BH:S mergers we have found. However, this is a subset of M1 where we have taken out the most resonant and hard encounters thus we should expect to find a larger increase in merger rates when dissipative terms are included. 

\subsection{Example interaction}
Here we introduce an example interaction where the change of flags changes the outcome of the encounter. The initial parameters are found in table \ref{tab:example_initialProperties} and the trajectories for the runs are found in figure \ref{fig:example_M2_traj}. For the no flags and PN runs (panel (a) and (b)) this interaction results in a flyby and the differences between the two runs are insignificant. When tides (panel (c)) are included, the interaction is more interesting; when the binary and single gets close, we form a new binary consisting of the star and the incoming BH with the third BH bound to this newly formed binary. After some time the binary and single have another close encounter where the star is exchanged out of the binary and kicked out of the interaction. The result is a BH-BH binary and an unbound star.

With PN + tides included the start of the encounter is very similar to the run with only tides; the star and the incoming BH forms a new binary with the other BH bound to them. However, at the second close passage we have another exchange of the star. The star is exchanged back to the initial binary BH and shortly after inspirals and merges with the BH.

In both the tides and the PN+tides runs we form a BH-BH binary after the interaction, we can thus look at the merger times due to gravitational wave radiation. We use equation \ref{eq:GWMergers} for this and find that for the tides run we have $a=8.15$ AU, $e=0.79$, $M_1=20.36$ M$_{\odot}$, $M_2 = 19.88$ M$_{\odot}$ which gives us $t_{GW} \approx 2.42 \cdot 10^{14}$ yrs. For the PN + tides run we use the sticky-star approximation to merge the star and the BH, we do not account for any mass loss and uses the centre of mass at the final time step as the new merger product. By looking at the orbit between the merger product and the third object we find that the orbit is bound with $a=6.27$ AU and $e=0.81$. The mass of the merger product is 21.01 M$_{\odot}$ while the mass of the other BH is 20.36 M$_{\odot}$. This gives a merger time of $t_{GW} \approx 5.82 \cdot 10^{13}$ yrs. The merger times of both of these binaries are very long, longer than a Hubble time. However, what we can see is that the binary that is created after the merger has a lower merger time. This binary can then encounter additional objects in the cluster and harden to the point where the BHs will merge.

\begin{table}
    \caption{Initial properties for example interaction extracted from M2 setup 1.}
    \begin{tabular}{ll}
        \toprule
        Property & Value \\
        \midrule
        M$_{BH,1}$ & 20.36 M$_{\odot}$   \\
        M$_{BH,2}$ & 19.88 M$_{\odot}$   \\
        M$_{\star}$ & 1.14 M$_{\odot}$ \\
        a & 1.13 AU \\
        e & 0.98 \\
        b &  14.01 AU \\
        V$_{\infty}$ & 26.58 km/s \\
        \bottomrule
    \end{tabular}
    \label{tab:example_initialProperties}
\end{table}

\section{Summary \& Conclusions}
We have performed binary-single scattering experiments using the \textsc{tsunami}  code with both manually setup interactions as well as interactions extracted from the \textsc{MOCCA}-SURVEY Database I. Each interaction was simulated four times using: (i) pure N-body, (ii) with energy dissipation due to 2.5PN terms, (iii) with energy dissipation due to tidal forces and (iv) with energy dissipation due to both PN terms and tidal forces. 

\begin{figure*}
    \centering
    \includegraphics[width=\textwidth]{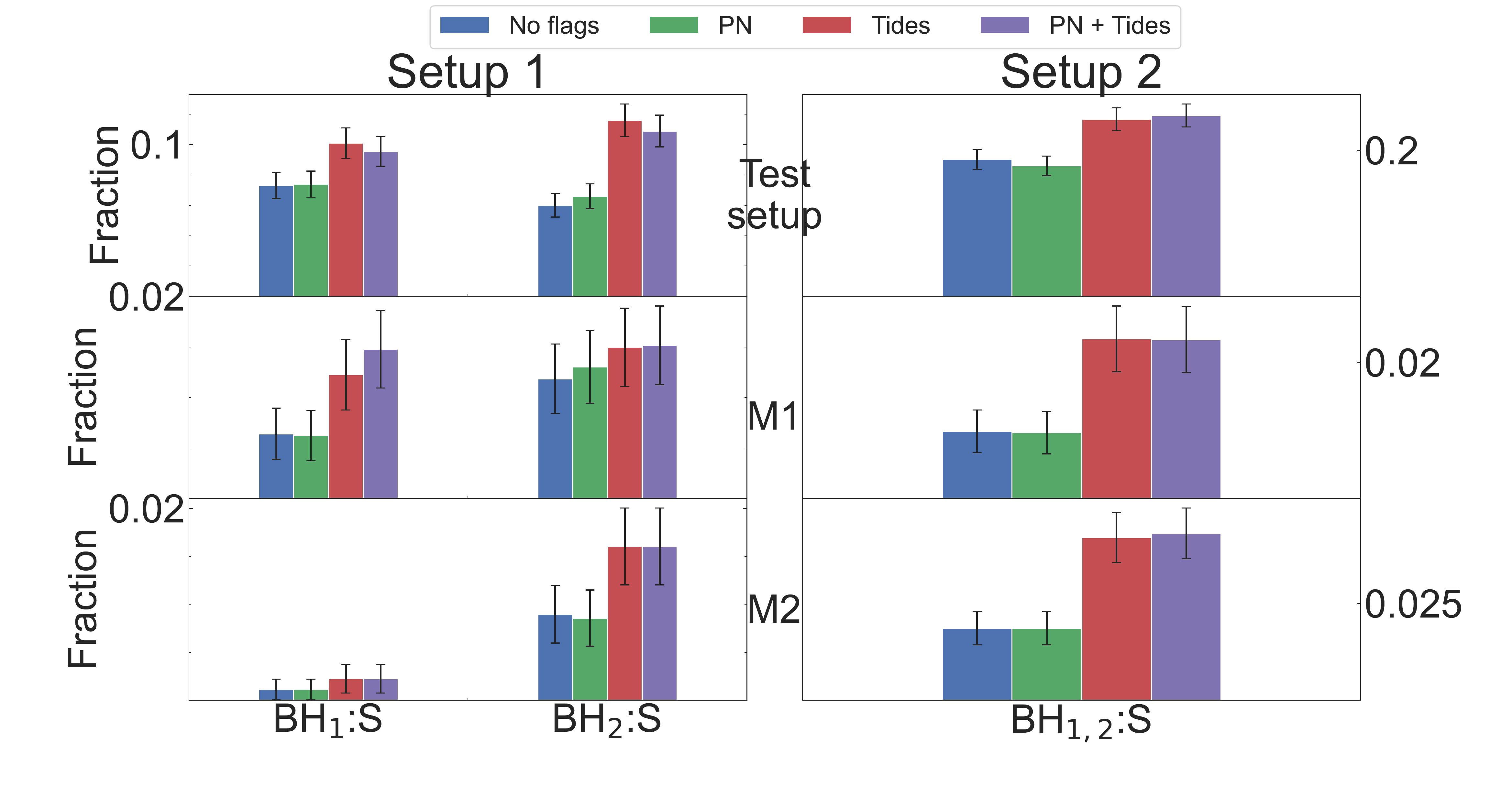}
    \caption{Histogram showing the fraction of BH:S mergers for both setups of the three data sets. The panels on the left show the BH$_1$:S and BH$_2$:S mergers for setup 1 and the panels on the right shows the BH$_{1,2}$:S mergers for setup 2. The error bars show the Poisson error distribution as $\pm \sqrt{N}$.}
    \label{fig:mergersHistoSummary}
\end{figure*}

We have found that the inclusion of both PN terms and tidal forces are important for binary-single interactions involving two BHs and a single star. Figure \ref{fig:mergersHistoSummary} shows the fraction of BH:S mergers for both setups of our three data sets relative to the total number of interactions for each set. These histograms show that the inclusion of tides increases the merger rates between BHs and stars during binary-single interactions by a significant amount for both setups. The magnitude of the increase is, in our data sets, dependent on the initial configuration of the data sets and varies significantly between our three sets.

The inclusion of PN terms increases the number of BH:BH mergers but does not affect the number of BH:star mergers, however, inclusion of tidal forces increase the number of BH-star mergers but do not affect the number of BH-BH mergers in a significant way. The magnitude of the increase in both BH:star and BH:BH mergers from the inclusion of tides and PN terms depends on the initial parameters of the interaction. 

The inclusion of PN terms increases the number of BH:BH mergers in all simulated data sets, in test setup 1 we get one merger with PN terms and zero without them. The runs without PN terms do not result in any BH:BH mergers in M1 setup 1 or 2, however, with PN terms included we see either two or three BH:BH mergers in each run.

When including tides in the case where we start with the star in the binary (setup 1) we see an increase in both the number of mergers between the incoming single BH and the star as well as an increase in the number of mergers between the star and the binary companion BH. As discussed above, the magnitude of the increase is different for our data sets; in our Test setup we find a 54 per cent increase in total number of mergers. For BH$_1$:S mergers we see an increase of 35 per cent and for BH$_2$:S mergers, an increase of 75 per cent. For MOCCA set 1 we see an increase in the number of total mergers of approximately 40 per cent while the number of BH$_1$:S mergers increase by approximately 100 per cent and BH$_2$:S mergers increase by approximately 26 per cent.

In MOCCA set 2 we can look at the total number of mergers as well as the number of unique mergers: with tides, we find an increase of approximately 70 per cent for the total number and 29 per cent for the number of unique mergers. The increase in the number of BH$_1$:S mergers is the same for the total number and number of unique mergers; approximately 90 per cent. The total number of B$_2$:S mergers increase by approximately 71 per cent and the number of unique mergers increase by 33 per cent. The reason behind the differences may be the masses of the objects and the initial setup of the interactions. In the test setup we see a larger increase in the number of BH$_2$:S mergers compared to the other sets. This is because most the binaries in M1 and M2 are and not as tight as the ones we created in the Test setup, thus it is harder for the incoming BH to perturb the orbit of the star to the point where it merges with the binary companion.

Tidal effects in the case where the initial setup is a BH-BH binary and a single star (setup 2) increase the number of BH-star mergers but the magnitude of the increase is, similarly to the case previously discussed, dependent on initial setup. For the test setup we see a 30 per cent increase in BH-star mergers with tidal effects compared to without. For M1 this increase is much larger at approximately 137 per cent. The increase is also larger for M2 at 125 per cent (total) and 110 per cent. Including PN terms in addition to tidal effects slightly increase the number of BH-star mergers in the test setup and MOCCA set 2 but decrease the number in MOCCA set 1.

The results summarized above show that there is a significant increase in the number of mergers between BHs and stars when tidal dissipation is included when computing the outcome of few-body interactions. This has important consequences on rates for transient events, like tidal disruptions caused by stellar-mass BHs within stellar clusters \citep{perets2016,kremer2019a, kremer2021}. Therefore, our findings indicate that the importance of including tidal dissipation for correctly computing the outcome of close encounters within stellar cluster simulations.

The inclusion of PN terms in addition to tides seems to have different effects on the number of BH-star mergers depending on which data set we use; for the test setup we see less BH-star mergers in the PN+tides run compared to the tides run. In MOCCA set 1 we have slightly more mergers between the incoming BH and the star with both PN+tides but no significant change in the number of mergers between the two initial binary components. For MOCCA set 2 we see no difference in the number of BH-star mergers between the PN+tides and the tides run.

The number of ionisations remain constant and we cannot see any significant difference between our runs that would indicate a dependence on additional processes. For individual interactions, the inclusion of additional processes may change the outcome to or from an ionisation but on a larger statistical scale there are no significant differences.

We found that the inclusion of the `impulsive' description of dynamical tides causes a very quick merger for a non-negligible number of binaries, to the point where the single does not interact with it. This only affects binaries from the M1 and M2 data set that had a very high eccentricity and thus a low pericenter distance. Since the eccentricity is set to 0 in the test setup, we do not see any of these very fast mergers, even though we have a very low initial semi-major axis for a few binaries. For M1 we find that 1.56 per cent of all interactions merge before the 3-body interaction, these quick mergers are found with the criteria described in section \ref{sec:instantMergers}. For M2 we find that in 4.1 per cent of all interactions, the binary merges before interacting with the single, however, if we only consider unique interactions (i.e. filter out identical interactions with different seed), we find that 0.9 per cent of binaries merge without the influence from the third object. The initial setup for M1 and M2 were taken from the \textsc{MOCCA}-Survey Database I simulations, in \textsc{MOCCA}, binary evolution is updated at the time of the interaction using prescriptions in the \textsc{BSE} code and then the \textsc{fewbody} code is called to compute the outcome of the interaction. Our results indicate that the treatment for tides in the \textsc{BSE} code do not capture tidally induced mergers of very eccentric binaries, where as when we use the dynamical treatment of tides with \textsc{tsunami} , we see that the binary components merge on very short timescales. This points towards the need to update the treatment for tides in binary population synthesis codes like \textsc{BSE}, as they may be underestimating the merger times for binaries with small pericenter distance values.

Many BH:S mergers leave behind a weakly bound BH binary with a very high eccentricity and low pericenter distance. It is possible that due to GW radiation during the pericenter passage that the BHs merge or that the orbit shrinks rapidly into an inspiral. However, since the orbit is very wide, the time it would take for this binary to complete one orbit is high and it is possible that a third object may disrupt the binary before this happens. It is possible to compare the interaction time scale in the clusters to the period of the binary to get an estimate of how many binaries would have time to complete one orbit before interacting again, however, estimating the interaction time scale is very difficult since this is very dependent on the cluster density, velocity dispersion, and where in the cluster the particular interaction takes place. Therefore, this paper puts more focus on the merger rates during the interactions and leave the possibility of following the binaries that are created as a result of the mergers for future studies.

Our results show an increase in mergers during binary-single interactions when additional processes are included. This increased merger rate may lead to a decrease in binary fraction which might influence overall cluster evolution. The inclusion of tidal effects are suspected to also have a large impact on close encounters when more than one star is involved. In interactions with two or more MS stars, the tidal effects may cause a merger between two MS stars which may in turn lead to a larger population of blue stragglers in the cluster. Additionally, tidal dissipation may also increase the number of disruption events involving interactions between BHs and evolved stars like giants \citep{ivanova2010,ivanova2017} and white dwarfs \citep{rosswog2009}. This can be an important channel for dynamically forming compact binary systems in dense stellar clusters. We will study this in future publications in which we plan on carrying out stellar cluster simulations with the \textsc{mocca} code in which \textsc{tsunami}  (with tidal dissipation and PN corrections) will be used to compute the outcome of close encounters between two and more bodies. 


\section*{Acknowledgements}
We would like to thank the reviewer for providing comments and suggestions that helped to improve the quality of the manuscript. LH, AA and MG were partially supported by the Polish National Science Center (NCN) through the grant UMO-2016/23/B/ST9/02732. LH, AA and RC acknowledge support from the Swedish
Research Council through the grant 2017-04217. AAT received support from JSPS KAKENHI Grant Numbers 17H06360, 19K03907 and 21K13914. JS is supported by the European Union’s Horizon 2020 research and innovation programme under Marie Sklodowska-Curie grant agreement No.
844629 and through Villum Fonden grant No. 29466.

\section*{Data Availability}

Input and output data for the scattering experiments carried out in this paper will be shared on request to the corresponding author. The \textsc{tsunami}  code and the simulated data will be shared on reasonable request to Alessandro A. Trani.



\bibliographystyle{mnras}
\bibliography{ref}

\bsp	
\label{lastpage}
\end{document}